\documentclass[journal, 9pt]{IEEEtran}
\ifCLASSINFOpdf
   \usepackage[pdftex]{graphicx}
  \graphicspath{{../pdf/}{../jpeg/}}
  \DeclareGraphicsExtensions{.pdf,.jpeg,.png}
\else
  \usepackage[dvips]{graphicx}
  \graphicspath{{../eps/}}
  \DeclareGraphicsExtensions{.eps}
\fi

\usepackage[cmex10]{amsmath}
\usepackage{amsthm}
\usepackage{amsfonts}
\usepackage{MnSymbol}

\usepackage{epstopdf}
\usepackage{graphicx}
\usepackage{array}
\usepackage{multirow}
\usepackage{hhline}
\usepackage{makecell}
\usepackage{pbox}
\usepackage[noadjust]{cite}
\usepackage{subfigure}
\usepackage{color}
\usepackage[export]{adjustbox}

\usepackage{bm}

\long\def\comment#1{}

\newfont{\bbb}{msbm10 scaled 700}

\newfont{\bb}{msbm10 scaled 1100}

\newcommand{\ev}{{\bf e}}
\newcommand{\fv}{{\bf f}}

\newcommand{\uv}{{\bf u}}

\newcommand{\vv}{{\bf v}}
\newcommand{\xv}{{\bf x}}

\newcommand{\zerov}{{\bf 0}}

\newcommand{\Bm}{{\bf B}}

\newcommand{\Dm}{{\bf D}}
\newcommand{\Em}{{\bf E}}
\newcommand{\Fm}{{\bf F}}
\newcommand{\Gm}{{\bf G}}

\newcommand{\Id}{{\bf I}}
\newcommand{\Jm}{{\bf J}}

\newcommand{\Lm}{{\bf L}}
\newcommand{\Mm}{{\bf M}}

\newcommand{\Pm}{{\bf P}}
\newcommand{\Qm}{{\bf Q}}

\newcommand{\Sm}{{\bf S}}

\newcommand{\Um}{{\bf U}}
\newcommand{\Wm}{{\bf W}}
\newcommand{\Vm}{{\bf V}}

\newcommand{\Cc}{{\cal C}}

\newcommand{\Ec}{{\cal E}}

\newcommand{\Gc}{{\cal G}}
\newcommand{\Hc}{{\cal H}}

\newcommand{\Oc}{{\cal O}}

\newcommand{\Sc}{{\cal S}}

\newcommand{\Vc}{{\cal V}}
\newcommand{\Xc}{{\cal X}}

\newcommand{\Lcb}{{\bm {\mathcal L}}}

\newcommand{\muv}{\hbox{\boldmath$\mu$}}

\newcommand{\Lambdam}{\hbox{\boldmath$\Lambda$}}

\newcommand{\Pim}{\hbox{\boldmath$\Pi$}}

\newcommand{\Thetam}{\hbox{\boldmath$\Theta$}}

\newcommand{\diag}{{\hbox{diag}}}

\newtheorem{theorem}{Theorem}

\newtheorem{lemma}{Lemma}
\newtheorem{definition}{Definition}

\usepackage{hyperref}
\usepackage{url}

\begin{document}
\title{Fast Graph Fourier Transforms Based on Graph Symmetry and Bipartition}
\author{Keng-Shih~Lu,~\IEEEmembership{Student Member,~IEEE},~and~Antonio~Ortega,~\IEEEmembership{Fellow,~IEEE}% 
\thanks{The authors are with the Department
of Electrical and Computer Engineering, University of Southern California, California,
CA 90089, USA (email: kengshil@usc.edu; ortega@sipi.usc.edu).}}% 
%\thanks{Manuscript received April 19, 2005; revised September 17, 2014.}}

%\markboth{Journal of \LaTeX\ Class Files,~Vol.~13, No.~9, September~2014}%
%{Shell \MakeLowercase{\textit{et al.}}: Bare Demo of IEEEtran.cls for Journals}


\maketitle

\begin{abstract}
The graph Fourier transform (GFT) is an important tool for graph signal processing, with applications ranging from graph-based image processing to spectral clustering. However, unlike the discrete Fourier transform, the GFT typically does not have a fast algorithm. In this work, we develop new approaches to accelerate the GFT computation. In particular, we show that Haar units (Givens rotations with angle $\pi/4$) can be used to reduce GFT computation cost when the graph is bipartite or satisfies certain symmetry properties based on node pairing. We also propose a graph decomposition method based on graph topological symmetry, which allows us to identify and exploit butterfly structures in stages. This method is particularly useful for graphs that are nearly regular or have some specific structures, e.g., line graphs, cycle graphs, grid graphs, and human skeletal graphs. Though butterfly stages based on graph topological symmetry cannot be used for general graphs, they are useful in applications, including video compression and human action analysis, where symmetric graphs, such as symmetric line graphs and human skeletal graphs, are used. Our proposed fast GFT implementations are shown to reduce computation costs significantly, in terms of both number of operations and empirical runtimes. 
\end{abstract}


\IEEEpeerreviewmaketitle

\section{Introduction}
%\IEEEPARstart{T}{his} demo file is intended to serve as a ``starter file''
%for IEEE journal papers produced under \LaTeX\ using
%IEEEtran.cls version 1.8a and later.
Graph signal processing (GSP) \cite{sandryhaila2013discrete,shuman2013emerging,ortega2018graph} is a framework that extends signal processing tools 
%from regularly sampled signals 
to data lying on irregular domains. In GSP, data points are represented as nodes in a graph, and relations between data points are captured by the graph edges. Data associated to the nodes is called a graph signal. 
Conventional signal processing tools such as the Fourier transform and filtering can be extended to signals defined on graphs, providing applications in sensor networks \cite{dong2013inference}, image and video processing \cite{cheung2018graph}, and machine learning \cite{chen2014semi-supervised}. 
% maybe cite G. Cheung, "Graph Spectral Image Processing" for image and video processing

As an extension of the discrete Fourier transform (DFT) to graph signals, the graph Fourier transform (GFT) is a fundamental tool in GSP. There are several definitions of the GFT \cite{shuman2013emerging,sandryhaila2013discrete,girault2018irregularity}, depending on whether the graph is directed, which graph shift operator is used (e.g. adjacency matrix or Laplacian matrix), and how the graph signal energy is defined. Following the definition in \cite{shuman2013emerging}, the GFT basis functions are defined as the eigenvectors of the graph Laplacian matrix, and their associated frequencies are the corresponding eigenvalues. The GFT coefficients of a given signal $\xv$ can be obtained by projecting $\xv$ onto the GFT basis functions. Those coefficients corresponding to smaller eigenvalues (lower frequency) reflect the energy of signal components with smaller variation on the graph. The GFT has a wide range of applications. 
First, based on the GFT and its frequency interpretation, graph spectral filters \cite{isufi2017autoregressive} can be defined by multiplying the GFT coefficients by the frequency response in the graph spectral domain, leading to applications such as image denoising and edge-preserving smoothing \cite{chen2014signal,onuki2016graph}. Second, when the graph signal is modeled by a Gaussian Markov random field (GMRF) \cite{rue2005gaussian}, the corresponding GFT can be regarded as the optimal decorrelating transform for that class of signals. 
Based on this fact, the GFT has been applied to image and video compression \cite{hu2015multiresolution,fracastoro2016graph,egilmez2015graph-based}. 
Third, in machine learning, when relations between data points are modeled by a graph, the GFT can be used for data clustering \cite{von_luxburg2007tutorial} and dimensionality reduction for classification \cite{menoret2017evaluating,tanaka2016dimensionality}. 

Unlike the DFT, which can be implemented with the well-known fast Fourier transform (FFT) algorithm \cite{cooley1965algorithm}, in general there are no fast algorithms to compute GFTs. 
%Fast algorithms are obtained for the DFT by exploiting the even or odd symmetry of its basis functions. 
%One can see that 
DFT basis functions are always even or odd symmetric, which can be exploited to obtain fast algorithms.
In contrast, arbitrary graph topologies do not always lead to Laplacian eigenvectors with such symmetry properties. Lack of fast algorithms is a significant drawback for GSP approaches, particularly when the GFT needs to be applied repeatedly. This has led researchers to investigate techniques for fast GFT computation (see Section \ref{subsec:relatedwork} for a  review of recent work). In particular, Magoarou et. al. have proposed a series of approaches for fast GFTs \cite{magoarou2016flexible,magoarou2016approximate,magoarou2018approximate}, which use optimization techniques to {\em approximate} a GFT 
by a fast transform 
constructed with a series of of parallel Givens rotations. 
%A more detailed review of those works can be founded in Section \ref{subsec:relatedwork}. 

Our work is motivated by noting that \emph{exact} fast GFTs 
are available for certain graphs with particular structures. 
For example, the discrete cosine transform (DCT) is known to be the GFT of a line graph with uniform edge weights \cite{strang1999discrete}, and it has well-known fast algorithms \cite{feig1992fast,kok1997fast}. Another example is a butterfly structured implementation for Type-4 DST \cite{han2013butterfly}, whose corresponding graph is a line graph with uniform weights with an added self-loop in the first node. 
%equal to 1, and a self-loop with a weight 2 added to the first node. 
Because of the availability of fast algorithms, DCT and Type-4 DST have been adopted in codecs such as HEVC \cite{sullivan2012overview} and AV1 \cite{aomedia}. Motivated by these fast algorithms, in this paper our goal is to explore more general classes of graphs with fast GFTs. In our preliminary work \cite{lu2016symmetric,lu2017fast}, we have introduced two classes of graphs, symmetric line and grid graphs, whose GFTs have a butterfly stage for fast implementation. 
This work extends and generalizes the results of \cite{lu2016symmetric,lu2017fast} to graphs that are bipartite or have more general symmetry properties. 
%In addition, we propose an approach for deriving fast GFTs for symmetric graphs, and evaluate the computation costs of these fast GFTs, in terms of number of operations  and empirical runtimes. 
%Similar to \cite{magoarou2016flexible,magoarou2016approximate,magoarou2018approximate}, this paper studies fast implementation methods for GFTs. However, unlike those methods, we explore the connection between graph structures and butterfly stages of the associated GFT, and derive {\em exact} fast GFTs rather than approximate ones. Under symmetry conditions, fast GFTs derived from our approach outperform existing methods in terms of both runtime and GFT accuracy. 
A more detailed outline of this paper will be presented in Section \ref{subsec:contributions}.

\subsection{Related Work}
\label{subsec:relatedwork}

\begin{figure}
    \centering
    \includegraphics[width=.45\textwidth]{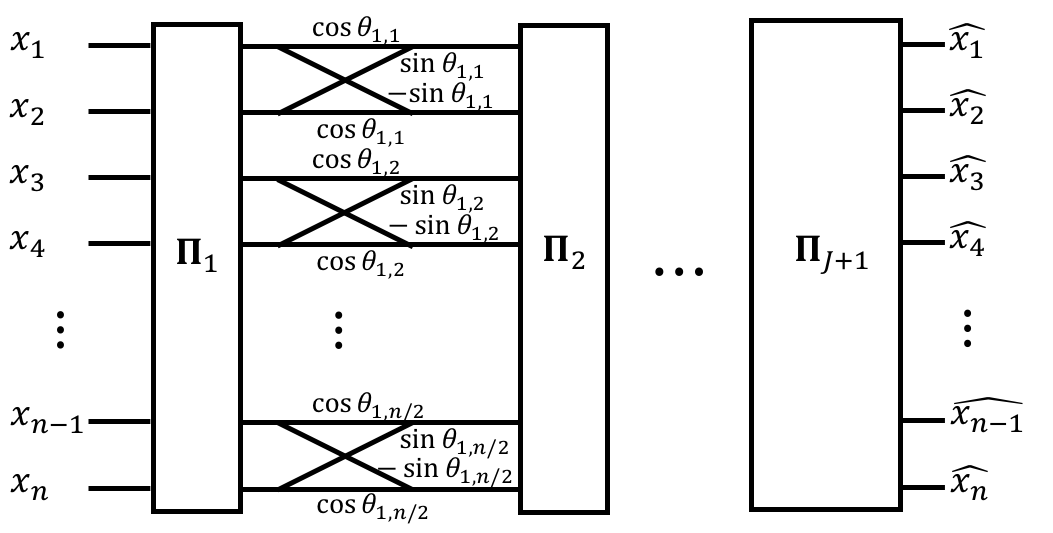}
    \caption{Fast transform using $J$ layers of Givens rotations. The parameter $0<\theta_{i,j}\leq\pi$ is the $j$-th rotation angle in the $i$-th butterfly stage, and $\Pim_k$ are permutation operations.}
    \label{fig:layeredgivens}
\end{figure}

An $n$ dimensional \emph{Givens rotation} \cite{golub1996matrix}, commonly referred to as a \emph{butterfly} \cite{han2013butterfly,magoarou2016approximate,magoarou2018approximate}, is a linear transformation that applies a rotation of angle $\theta$ to two coordinates, denoted as $p$ and $q$. Its associated matrix $\Thetam(p,q,\theta)$ has the form: 
\begin{equation}
\label{eq:givens}
  \left\{ \begin{array}{ll}
  \Theta_{pp}=\Theta_{qq}=\cos\theta, \\
  \Theta_{qp}=-\Theta_{pq}=\sin\theta, \\
  \Theta_{ii}=1, & i\neq p,\; q, \\
  \Theta_{ij}=0, & \text{otherwise}.
  \end{array} \right.
\end{equation}
A system using layers of parallel Givens rotations (e.g.,  Fig.~\ref{fig:layeredgivens}), can be used to design a fast approximate transform. In particular, each Givens rotation can be implemented using three lifting steps \cite{daubechies1998factoring}, which further reduces the number of operations involved.
%The choice of Givens rotations involves the selection of node pairs ($p$ and $q$) and angles ($\theta$). These node pairs can be selected using greedy-based pairing strategy \cite{chen2011design} or matching algorithm on graph \cite{li2017layered}, while the optimal rotation angles can be determined using Jacobi's method \cite{golub1996matrix} once the node pairs are fixed. 

Recently, several papers have focused on GFT-specific speedup techniques. The work in \cite{magoarou2016flexible} uses a gradient-descent-based optimization approach to approximate the GFT matrix by a product of sparse matrices, while \cite{magoarou2016approximate} refines this method such that the resulting transform matrix can approximately diagonalize the graph Laplacian. In \cite{magoarou2018approximate}, a truncated Jacobi algorithm was introduced for picking the Givens rotations used in the approximate fast GFT, leading to an implementation with the structure shown in Fig.~\ref{fig:layeredgivens}. This approach was further analyzed in \cite{magoarou2017analyzing}, which demonstrates that more Givens rotations are required to approximate the Laplacian eigenvectors whose corresponding eigenvalues are close. Although these methods \cite{magoarou2016flexible,magoarou2016approximate,magoarou2018approximate} are able to find approximate fast GFTs, they do so without taking advantage of structural properties of the original graph. 

% \KS{mention the methods for learning fast transforms \cite{lu2017graph,cao2011sparse}}
% \KS{Include the references 19-24 in \cite{magoarou2018approximate}}
% \cite{cao2011sparse}

\subsection{Contributions}
\label{subsec:contributions}

The relation of topological properties of graphs, such as bipartition, repeated subgraphs, symmetry, and uniformity of weights, to the structure of the GFT bases is an important topic in GSP. In this work, we show that for graphs with certain symmetry or bipartition properties, {\em exact and fast} GFTs based on \emph{Haar units} (butterflies with rotation angle $\pi/4$) can be designed. We propose divide-and-conquer fast GFT algorithms for symmetric graphs and demonstrate that the resulting fast GFTs lead to significant complexity reduction, potentially beneficial  in hardware implementation or in scenarios where the graph is fixed and the corresponding GFT is applied multiple times. We show that graphs for which such Haar-unit-based fast GFTs can be developed are useful in applications such as video coding and human activity analysis.  

Unlike fast approximate GFTs  \cite{magoarou2016flexible,magoarou2016approximate,magoarou2018approximate}, our fast GFTs are based on graph topological properties, and are exact. Experimental results show that as long as the desired graph symmetry property is available, our fast GFTs can provide outperform the approach in \cite{magoarou2018approximate} in terms of speed.
%, with zero or smaller approximation error. 
With respect to our earlier work \cite{lu2016symmetric,lu2017fast}, the main novelties of this paper are: 1) we define a notion of graph symmetry that gives rise to butterfly implementation of the GFT, and show that the results in \cite{lu2016symmetric,lu2017fast} are particular cases within this general framework; 2) we introduce, in addition to line and grid graphs, more examples of graphs with fast GFTs, such as star graphs, cycle graphs, and skeleton graphs; 3) we provide more comprehensive results, including experimental runtimes and comparisons with existing approaches. 

The rest of this paper is organized as follows. Section \ref{sec:preliminaries} introduces notation and basic graph signal processing concepts. In Section \ref{sec:haar} we derive the algebraic conditions for a GFT to have a left or right butterfly stage. In Section \ref{sec:sym_graph} we define graph symmetry based on node pairings, and propose a graph decomposition method for designing fast GFTs. In Section \ref{sec:examples} we show several examples of fast GFTs based on the proposed method, and highlight some applications of the derived fast GFTs. Section \ref{sec:exp} provides experimental results to demonstrate the runtime reduction provided by the fast GFTs. Finally, Section \ref{sec:conclusion} concludes this paper. 

\section{Preliminaries}
\label{sec:preliminaries}

\subsection{Notations and Conventions}
We use bold symbols to denote vectors and matrices. The $n\times n$ identity matrix is denoted by $\Id_n$. The $n\times n$ order-reversal permutation matrix is denoted by 
\begin{equation}
\label{eq:jmat}
  \Jm_n=\begin{pmatrix}
&&&1\\&&1&\\&\udots&&\\1&&&
\end{pmatrix}.
\end{equation}
When $\Jm_n$ right (resp.~left) multiplies another matrix, it flips this matrix left to right (resp.~up to down). In \eqref{eq:jmat} and in what follows, the entries not included in the matrix are meant to be zero. The subscripts of $\Id$ and $\Jm$ matrices indicate their sizes, and may be omitted for brevity. 
For scalars $a_1,\dots,a_k$ and square matrices $\Mm_i$ with arbitrary sizes, we denote diagonal and block diagonal matrices in compact notation as
\begin{align*}
  & \text{diag}(a_1,a_2,\dots,a_k)=\begin{pmatrix}
    a_1 &&& \\ &a_2&& \\ && \ddots &\\ &&&a_k \end{pmatrix}, \\
  & \text{diag}(\Mm_1,\Mm_2,\dots,\Mm_k)=\begin{pmatrix}
  \Mm_1 & & & \\
   & \Mm_2 &  & \\
   & & \ddots & \\
   & & & \Mm_k
  \end{pmatrix}.
\end{align*}
Finally, the set of $n$-dimensional real-valued vectors is denoted as $\mathbb{R}^n$, and the set of $n\times n$ orthogonal matrices (with columns normalized to have unit norms) is denoted as $\mathbb{O}^n$. 

\subsection{Graph Fourier Transform}
%In graph signal processing \cite{shuman2013emerging,sandryhaila2013discrete}, 
In this paper, we focus on undirected graphs\footnote{We leave the extension to directed graphs for future work.}. Let $\xv$ be a length-$n$ graph signal associated to an undirected graph $\Gc(\Vc,\Ec,\Wm)$. In particular, there are $n$ nodes in the vertex set $\Vc$, each corresponding to an element of $\xv$. Each edge $e_{ij}\in\Ec$ describes the inter-sample relation between nodes $i$ and $j$. $\Wm$ is the weighted adjacency matrix, whose $(i,j)$ entry, $w_{i,j}$, is the weight of the edge between nodes $i$ and $j$, and $s_i:=w_{ii}$ is the weight of the self-loop on node $i$. The (unnormalized) graph Laplacian matrix of $\Gc$ is: 
%defined as 
\begin{equation}\label{eq:def_laplacian}
\Lm=\Dm-\Wm+\Sm, 
\end{equation}
where $\Sm=\text{diag}(s_1,\dots,s_n)$ is the diagonal self-loop matrix, and the degree matrix $\Dm=\text{diag}(d_1,\dots,d_n)$ is a diagonal matrix with $d_i=\sum_{j=1}^n w_{i,j}$. A graph is \emph{bipartite} if its vertices can be divided into two disjoint sets (or, two \emph{parts}) $\Sc_1$ and $\Sc_2$ such that every edge connects a vertex in $\Sc_1$ and one in $\Sc_2$. 

By definition of graph Laplacian \eqref{eq:def_laplacian}, one can express the self-loop and edge weights in terms of entries of the Laplacian matrix $\Lm=(l_{i,j})_{i,j}$, and vice versa:
\begin{align}
\label{eq:sw_from_l}
  & s_i=\sum_{j=1}^n l_{i,j},\quad w_{i,j}=-l_{i,j}\text{ for } i\neq j, \\
\label{eq:l_from_sw}
  & l_{i,i} = s_i + \sum_{\substack{j=1 \\ j\neq i}}^n w_{i,j},\quad l_{i,j}=-w_{i,j}\text{ for } i\neq j.
\end{align}

The Graph Fourier Transform (GFT), also known as Graph-Based Transform (GBT), is obtained from the eigen-decomposition of the graph Laplacian matrix, $\Lm=\Um\Lambdam\Um^\top$, where $\Um$ is the matrix of eigenvectors and $\Lambdam$ is the diagonal matrix of eigenvalues. The $i$-th coefficient of GFT of a graph signal $\xv$ is defined as the projection of $\xv$ onto $\uv_i$, the $i$-th column of $\Um$. In some applications, such as spectral clustering \cite{von_luxburg2007tutorial}, it may be beneficial to use a GFT defined on the eigenvectors of the \emph{symmetric normalized  Laplacian} $\Lcb=\Dm^{-1/2}\Lm\Dm^{-1/2}$. In what follows, we use GFTs associated to the unnormalized Laplacian matrix, unless stated otherwise.

GFT coefficients provide a frequency representation of the given signal, since GFT basis functions associated to lower (resp. higher) eigenvalues represent lower (resp. higher) variation on the graph. To see this, we note that the Laplacian quadratic form 
\begin{equation}
\label{eq:lqf}
  \fv^\top\Lm\fv=\sum_{(i,j)\in\Ec}w_{i,j}(f_i-f_j)^2+\sum_{k=1}^n s_k f_k^2
\end{equation}
measures the variation of signal $\fv$ on the graph. Since $w_{i,j}$ and $s_k$ are non-negative, $\Lm$ is positive semi-definite and thus $\fv^\top\Lm\fv$ is always non-negative. The eigenvectors of $\Lm$ are the solutions to 
\begin{align*}
  \uv_1 = \underset{\|\fv\|=1}{\text{argmin}} \quad \fv^\top\Lm\fv,\quad
  \uv_{k} = \underset{\fv\perp \uv_1,\dots,\uv_{k-1},\|\fv\|=1} {\text{argmin}} \quad \fv^\top\Lm\fv.
\end{align*}
Thus, eigenvectors $\uv_1$, $\dots$, $\uv_n$ form an orthogonal basis with functions having lower to higher variations on the graph. The quantities of their corresponding variations are given by the associated eigenvalues $\lambda_1$, $\dots$, $\lambda_n$, which are also called \emph{graph frequencies}.

\section{Algebraic Conditions for Haar Units in GFTs}
\label{sec:haar}

\begin{figure}
    \centering
    \includegraphics[width=.20\textwidth]{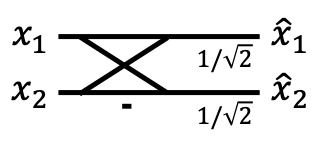}
    \caption{The Haar unit.}
    \label{fig:haar}
\end{figure}

\begin{figure}
\centering
\subfigure[]{
\includegraphics[width=.23\textwidth]{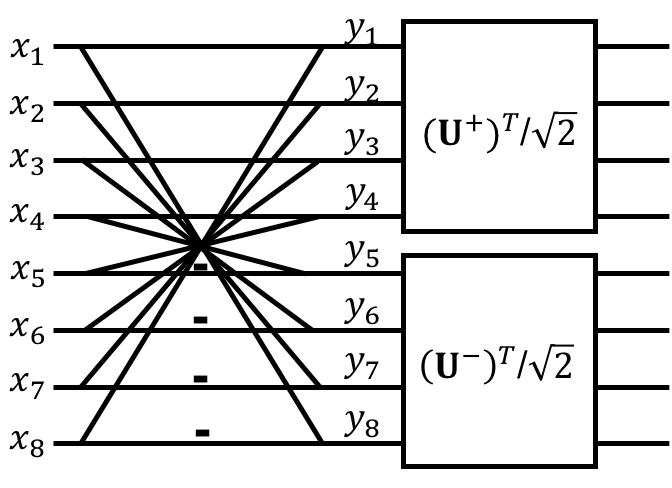}}
\subfigure[]{
\includegraphics[width=.23\textwidth]{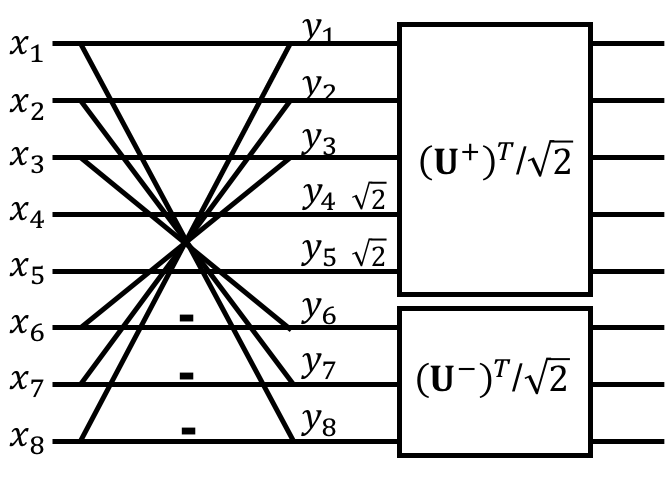}}
\subfigure[]{
\includegraphics[width=.23\textwidth]{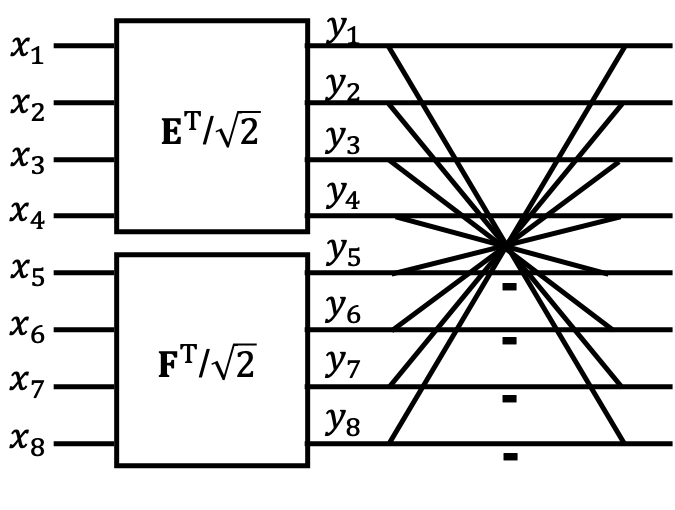}}
\subfigure[]{
\includegraphics[width=.23\textwidth]{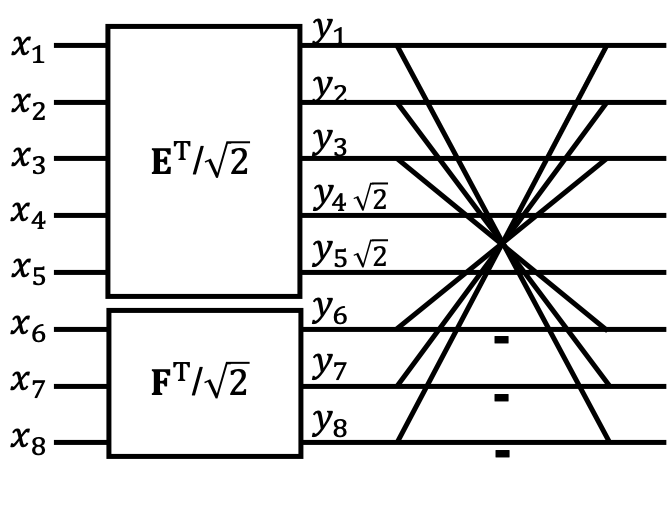}}
\caption{Examples of fast algorithms using butterfly stages with $n=8$ . (a)(b) Left butterfly stages. (c)(d) Right butterfly stages.}
\label{fig:diagram}
\end{figure}

In this paper, we improve computation efficiency by using the elementary operation in Fig.~\ref{fig:haar}, which is equivalent to a 2$\times$2 Haar transform. This operation is a Givens rotation with angle $\pi/4$, followed by a sign flip. In particular, in Fig.~\ref{fig:haar}, 
\[ 
\begin{pmatrix}\hat{x}_1\\\hat{x}_2\end{pmatrix}=\begin{pmatrix}1&0\\0&-1\end{pmatrix}\cdot\Theta^\top(1,2,\pi/4)\cdot\begin{pmatrix}x_1\\x_2\end{pmatrix}. 
\] 
We refer to the operator in Fig.~\ref{fig:haar} as \emph{Haar unit}, as opposed to general Givens rotations, which are often referred to as ``butterflies'' \cite{han2013butterfly,li2017layered,magoarou2018approximate}. 
We say that a \emph{butterfly stage} is a stage in a transform diagram with several parallel Givens rotations or Haar units. For example, in Fig.~\ref{fig:diagram}(a), we call the stage that produces $y_i$ from $x_i$ a butterfly stage, and the operator that produces $y_1$ and $y_8$ from $x_1$ and $x_8$ a Haar unit of this butterfly stage. Note that the factor $1/\sqrt{2}$ of the Haar unit can usually be absorbed into other stages of the transform computation (see Fig.~\ref{fig:diagram}(a) as an example, where the factor $1/\sqrt{2}$ is merged into the later stage). Thus, a Haar unit typically requires an addition and a subtraction only.

\begin{figure}
\centering
\subfigure[]{
\includegraphics[width=.17\textwidth]{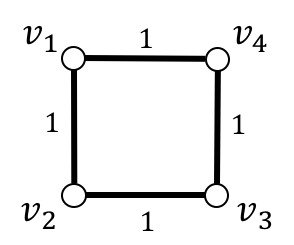}}
\subfigure[]{
\includegraphics[width=.25\textwidth]{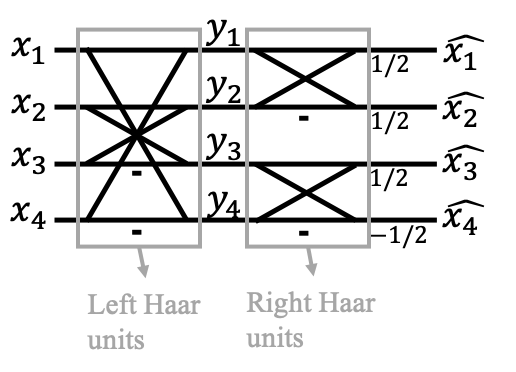}}
\caption{(a) The 4-node cycle graph and (b) a fast algorithm for its GFT.}
\label{fig:diagram_c4}
\end{figure}

We consider a divide and conquer framework based on stages of Haar units and parallel sub-transforms, as illustrated Fig.~\ref{fig:diagram}. For each Haar unit, we always assume that the two output variables, such as $y_1$ and $y_8$ in Fig.~\ref{fig:diagram}(a), will be inputs of different sub-transforms in the \emph{next} stage. Otherwise, such a Haar unit, e.g., the one acting on $x_1$ and $x_8$, can be trivially absorbed into the next stage.  

As a first example, we consider a 4-node cycle graph with no self-loops and unity edge weights as in Fig.~\ref{fig:diagram_c4}(a). It has a GFT matrix
% \[
%   \Um_{\Cc_4}=\begin{pmatrix}
%   1/2 & 1/\sqrt{2} & 0 & 1/2 \\ 
%   1/2 & 0 & -1/\sqrt{2} & -1/2 \\
%   1/2 & -1/\sqrt{2} & 0 & 1/2 \\ 
%   1/2 & 0 & 1/\sqrt{2} & -1/2 \\
%   \end{pmatrix}.
% \] 
\[
  \Um_{\Cc_4}=\frac{1}{2}\begin{pmatrix}
  1 & 1 & 1 & 1 \\ 
  1 & -1 & 1 & -1 \\
  1 & -1 & -1 & 1 \\ 
  1 & 1 & -1 & -1 \\
  \end{pmatrix}.
\] 
Based on the structure of $\Um_{\Cc_4}$, it can be seen that GFT can be implemented using two butterfly stages, as in Fig.~\ref{fig:diagram_c4}(b). In what follows, we refer to first-stage Haar units  (e.g., the one acting on $x_1$ and $x_4$) as \emph{left Haar units}, and to those in the last stage (such as the one producing $\widehat{x_1}$ and $\widehat{x_2}$) as \emph{right Haar units}. We will explore the conditions that allow a GFT to be factored into terms that include left and right Haar units, which will enable us to develop  techniques for designing such fast GFTs. We will show that \emph{right Haar units are associated to bipartite graphs} (Section \ref{subsec:right_pm}), while \emph{left Haar units are related to graph symmetries}  (Section \ref{subsec:left_pm}). 

Note that the second and third columns of $\Um_{\Cc_4}$ correspond to eigenvalue 2 (which has multiplicity 2). This means that the GFT basis is not unique, because we can obtain another orthogonal basis for the eigenspace corresponding to eigenvalue 2. An example of another basis for this GFT is the length-4 DFT, which has a well-known fast algorithm \cite{cooley1965algorithm}. Despite this non-uniqueness, a fast algorithm for a particular GFT basis would still be useful: we can first apply it to obtain coefficients for this particular basis, then apply an $m$-dimensional rotation ($m\times m$ orthogonal transform) on those $m$ GFT coefficients associated to eigenvalues with a multiplicity $m>1$, to obtain coefficients associated to another GFT basis. For example, if we properly apply a rotation to $\widehat{x_2}$ and $\widehat{x_3}$ in Fig.~\ref{fig:diagram_c4}(b), we can obtain the second and third DFT coefficients. In what follows, we study GFT implementations for which stages of Haar units are available. In cases where eigenvalues of high multiplicity are present, we favor the set of eigenvectors for the corresponding subspace that will lead to a more efficient implementation. 

For a general graph with $n$ nodes, we define the following $n\times n$ orthogonal matrix to represent a stage of $p$ parallel Haar units (with $p\leq n/2$):
\begin{equation}
\label{eq:Bnp}
  \Bm_{n,p}=\frac{1}{\sqrt{2}}
  \begin{pmatrix}
    \Id_p & \zerov & \Jm_p \\ \zerov & \sqrt{2}\Id_{n-2p} & \zerov \\ \Jm_p & \zerov & -\Id_p 
  \end{pmatrix}.
\end{equation}
Note that $\Bm_{n,p}^\top=\Bm_{n,p}$, and when we multiply a vector $\xv=(x_1,\dots,x_n)^\top$ by $\Bm_{n,p}$, we have
\[
  (\Bm_{n,p}\cdot\xv)_i=\left\{\begin{array}{ll} \frac{1}{\sqrt{2}}(x_i+x_{n+1-i}), & i=1,\dots,p \\
  x_i, & i=p+1,\dots,n-p \\
  \frac{1}{\sqrt{2}}(-x_i+x_{n+1-i}), & i=n-p+1,\dots,n
  \end{array}\right.
\]
For example, $\Bm_{8,4}$ are $\Bm_{8,3}$ are equivalent to the butterfly stages in Figs.~\ref{fig:diagram}(a) and (b), respectively, with a scaling constant $1/\sqrt{2}$. 
The factors $\sqrt{2}$ and $1/\sqrt{2}$ are included in \eqref{eq:Bnp} so that the columns of $\Bm_{n,p}$ have unit norms; in this way, when $\Um$ is an orthogonal matrix and $\Um=\Bm_{n,p}\bar{\Um}$, then $\bar{\Um}$ is orthogonal as well, meaning that $\Um$ can be factorized into a butterfly stage and another orthogonal transform.

\subsection{Conditions for Right Haar Units}
\label{subsec:right_pm}

Let an orthogonal transform $\Um$ have a right butterfly stage with $p$ Haar units, and assume without loss of generality that the entries of input and output vectors are properly ordered. Then, in compact notation, the GFT of input $\xv$ 
can be written as
\begin{equation}
\label{eq:right_gft}
\Um^\top\xv=\Bm_{n,p}\cdot\diag(\Em^\top,\Fm^\top)\cdot\xv,
\end{equation}
where $\Em\in \mathbb{O}^{n-p}$ and $\Fm\in \mathbb{O}^{p}$. This means that
\begin{align}
\label{eq:right_pm}
  \Um &= \begin{pmatrix}
    \Em_{11} & \Em_{12} & \zerov \\ \Em_{21} & \Em_{22} & \zerov \\
    \zerov & \zerov & \Fm
  \end{pmatrix}\frac{1}{\sqrt{2}}
  \begin{pmatrix}
    \Id_p & \zerov & \Jm_p \\ \zerov & \sqrt{2}\Id_{n-2p} & \zerov \\
    \Jm_p & \zerov & -\Id_p
  \end{pmatrix} \nonumber\\
  &= \frac{1}{\sqrt{2}}
  \begin{pmatrix}
    \Em_{11} & \sqrt{2}\Em_{12} & \Em_{11}\Jm_p \\ 
    \Em_{21} & \sqrt{2}\Em_{22} & \Em_{21}\Jm_p \\
    \Fm\Jm_p & \zerov & -\Fm
  \end{pmatrix},
\end{align}
where $\Em_{11}$, $\Em_{12}$, $\Em_{21}$, and $\Em_{22}$ are subblock components of $\Em$.
Recall that $\Jm$ flips a matrix left to right when right-multiplied. 
Thus, for $k=1,\dots,p$, if we denote the $k$-th column of $\Um$ as $\uv_k=(\ev_k^\top, \fv_{p-k+1}^\top)^\top$, then the $(n-k+1)$-th column of $\Um$ is $(\ev_k^\top, -\fv_{p-k+1}^\top)^\top$. GFT matrices with this structure 
arise from k-regular bipartite graphs (k-RBGs):
% \cite{narang2011downsampling}
\begin{lemma}[\cite{jakobson1999eigenvalue}]
\label{lem:k-rbg}
Let $\Lm$ be the Laplacian of a k-RBG with $\Sc_1=\{1,\dots,n/2\}$ and $\Sc_2=\{n/2+1,\dots,n\}$. If $\uv=(\uv_1^\top,\; \uv_2^\top)^\top$ with $\uv_1,\uv_2\in\mathbb{R}^{n/2}$ is an eigenvector of $\Lm$ with eigenvalue $\lambda$, then $\hat{\uv}=(\uv_1^\top,\; -\uv_2^\top)^\top$ is an eigenvector of $\Lm$ with eigenvalue $2k-\lambda$.
\end{lemma}

In this lemma, we fix $p=n-p=n/2$, which follows from the k-RBG topology. Although Lemma 1 was introduced in \cite{jakobson1999eigenvalue}, where only unweighted graphs are considered, it can be trivially generalized for weighted graphs. Lemma~\ref{lem:k-rbg} also leads to the following theorem.
\begin{theorem}
\label{thm:right_haar}
Let $\Lm$ be the Laplacian of a k-RBG with $\Sc_1=\{1,\dots,n/2\}$ and $\Sc_2=\{n/2+1,\dots,n\}$, then there exists a GFT matrix that has the structure \eqref{eq:right_pm}. Therefore, this GFT has a stage of right Haar units. 
\end{theorem}
\noindent {\it Proof:} We prove this theorem by construction. Since the graph Laplacian is symmetric, it has $n$ linearly independent eigenvectors, which enables us to construct a set $\Hc$ of $n$ eigenvectors as follows.
\begin{itemize}
    \item[a.] $\Hc\leftarrow\phi$.
    \item[b.] Pick an eigenvector $\muv=(\muv_1^\top;\muv_2^\top)^\top$ of $\Lm$ that is not in the span of $\Hc$. Note that this vector is guaranteed to exist as long as $\Hc$ has less than $n$ elements.
    \item[c.] If $\muv_2=\zerov$, then let $\Hc\leftarrow\Hc\cup\{\muv\}$. Otherwise, let $\muv'=(\muv_1^\top;-\muv_2^\top)^\top$, which, by Lemma~\ref{lem:k-rbg}, is also an eigenvector of $\Lm$ and does not belong to the span of $\Hc$. Then, we set $\Hc\leftarrow\Hc\cup\{\muv,\muv'\}$.
    \item[d.] Repeat b. and c. until $\Hc$ has $n$ elements. \qed
\end{itemize}
Theorem~\ref{thm:right_haar} provides certain sufficient (but, in fact, not necessary) conditions for a GFT to have a stage of right Haar units, even though the GFT matrix may not be unique. The matrices $\Em$ and $\Fm$ can be obtained by right multiplying the target GFT matrix $\Um$ by $\Bm_{n,p}$: 
\[
  \diag(\Em,\Fm)=\Um\cdot\Bm_{n,p}.
\]

If we consider eigenvectors of a \emph{normalized} Laplacian, a result similar to Theorem~\ref{thm:right_haar} can be derived.
% \begin{lemma}[\cite{chung1997spectral}]
% \label{lem:norm-bg}
% Let $\Lcb$ be the normalized Laplacian of a bipartite graph with $\Sc_1=\{1,\dots,p\}$ and $\Sc_2=\{p+1,\dots,n\}$. If $\uv=(\uv_1^\top,\; \uv_2^\top)^\top$ with $\uv_1\in\mathbb{R}^{p}$ and $\uv_2\in\mathbb{R}^{n-p}$ is an eigenvector of $\Lcb$ with eigenvalue $\lambda$, then $\hat{\uv}=(\uv_1^\top,\; -\uv_2^\top)^\top$ is an eigenvector of $\Lcb$ with eigenvalue $2-\lambda$.
% \end{lemma}
\begin{theorem}
Let $\Lcb$ be the normalized Laplacian of a bipartite graph with $\Sc_1=\{1,\dots,p\}$ and $\Sc_2=\{p+1,\dots,n\}$, then there exists a GFT matrix that has the structure \eqref{eq:right_pm}.
\end{theorem}
We omit the proof for brevity. In this case, $\Gc$ only needs to be bipartite (rather than k-regular bipartite), and $p$ need not be $n/2$.

\subsection{Conditions for Left Haar Units}
\label{subsec:left_pm}

\begingroup
\renewcommand*{\arraystretch}{1.3}
\begin{table}[t]
\centering
\caption{Definitions of symmetries for vectors, matrices, and graphs.}
\label{tab:syms}
\begin{tabular}{|l|l|l|}
\hhline{|=|=|=|}
Subject & Terminology & Definition \\
\hhline{|=|=|=|}
\multirow{2}{*}{Vector $\vv$} & Even symmetric & $\vv=\Jm\vv$ \\
& Odd symmetric & $\vv=-\Jm\vv$ \\
\hline
\multirow{3}{*}{Matrix $\Mm$} & Symmetric & $\Mm=\Mm^\top$ \\
& Centrosymmetric \cite{cantoni1976eigenvalues} & $\Mm=\Jm\Mm^\top\Jm$ \\
& Bisymmetric \cite{cantoni1976eigenvalues} & $\Mm=\Mm^\top=\Jm\Mm^\top\Jm$ \\
\hline
Graph & \multirow{2}{*}{$\phi$-symmetric (Definition \ref{def:graph_sym})} & \multirow{2}{*}{$w_{i,j}=w_{\phi(i),\phi(j)},\;\forall i,j$} \\
$\Gc(\Vc,\Ec,\Wm)$ & & \\
\hhline{|=|=|=|}
\end{tabular}
\end{table}
\endgroup

If $n$ is even and the GFT has a butterfly stage in the left with exactly $n/2$ Haar units as in Fig.~\ref{fig:diagram}(a), then 
\begin{equation}
\label{eq:umat_left_haar}
  \Um=\Bm_{n,n/2}
  \begin{pmatrix}
    \Um^+ & \zerov \\ \zerov & \Um^-
  \end{pmatrix}
  =\frac{1}{\sqrt{2}}\begin{pmatrix}
    \Um^+ & \Jm\Um^- \\ \Jm\Um^+ & -\Um^-
  \end{pmatrix},
\end{equation}
where $\Um^+, \Um^- \in\mathbb{O}^{n/2}$ denote non-zero block components that characterize two sub-transforms as in Fig.~\ref{fig:diagram}(a).\footnote{The symbols $\Um^+$ and $\Um^-$ are chosen for consistency with Haar units. They are used to denote sub-GFTs, as will become clear in Sec.~\ref{subsec:decomposition}.} From the right hand side of \eqref{eq:umat_left_haar} we see that each column $\uv_i$ of $\Um$ must be either \emph{even symmetric} (i.e., $\uv_i=\Jm\uv_i$) or \emph{odd symmetric} (i.e., $\uv_i=-\Jm\uv_i$). In this case, the Laplacian must be \emph{centrosymmetric} (symmetric around the center):
% \cite{andrew1973eigenvectors}
\begin{lemma}[\cite{cantoni1976eigenvalues}]
\label{lem:centrosym}
Let $n$ be even. An $n\times n$ matrix $\Qm$ has a set of $n$ linearly independent eigenvectors that are even or odd symmetric if and only if $\Qm$ is centrosymmetric, i.e., %$q_{ij}=q_{N-j+1,N-i+1}$, or, 
$\Qm=\Jm\Qm^\top\Jm$. 
\end{lemma}
Note that the Laplacian matrix $\Lm$ of an undirected graph is always symmetric ($\Lm=\Lm^\top$), but an additional centrosymmetry condition ($\Lm=\Jm\Lm^\top\Jm$) is required so that Lemma \ref{lem:centrosym} holds. Such a matrix with both symmetries ($\Lm=\Lm^\top=\Jm\Lm^\top\Jm$) is called \emph{bisymmetric}, and its entries are symmetric around both diagonals. 
%For clarity, definitions related to symmetry for vectors, matrices, and graph topology (as will be discussed later), 
The various types of symmetries considered in this paper listed in Table \ref{tab:syms}. 

Lemma~\ref{lem:centrosym} states that for even $n$, a GFT can be factored to include $n/2$ left Haar units if and only if the associated Laplacian matrix is bisymmetric (with nodes properly ordered). 
We now generalize this result to the case when there are only $p<n/2$ Haar units in the first butterfly stage, and with a possibly odd $n$. Again, we assume without loss of generality that the graph nodes, input, and output variables are properly ordered (notations defined for general node ordering will be introduced in Section \ref{sec:sym_graph}). We let $\Vc_X=\{1,\dots,p\}$, $\Vc_Z=\{p+1,\dots,n-p\}$, and $\Vc_Y=\{n-p+1,\dots,n\}$ be disjoint subsets of vertices. We define
\begin{align}
\label{eq:BLB}
  \Gm &:= \Bm_{n,p}^\top \cdot \Lm \cdot \Bm_{n,p},
  %&= \diag(\Em,\Fm) \cdot \Dm \cdot \diag(\Em^\top,\Fm^\top),
\end{align}
and denote the corresponding subblock components of $\Lm$ and $\Gm$ as
\begin{equation}
\label{eq:L_123}
  \Lm=\begin{pmatrix}
  \Lm_{XX} & \Lm_{XZ} & \Lm_{XY} \\ 
  \Lm_{ZX} & \Lm_{ZZ} & \Lm_{ZY} \\
  \Lm_{YX} & \Lm_{YZ} & \Lm_{YY}
  \end{pmatrix},\; \Gm = \begin{pmatrix}
  \Gm_{XX} & \Gm_{XZ} & \Gm_{XY} \\ 
  \Gm_{ZX} & \Gm_{ZZ} & \Gm_{ZY} \\
  \Gm_{YX} & \Gm_{YZ} & \Gm_{YY}
  \end{pmatrix}.
\end{equation} 
Similar to \eqref{eq:right_gft} and \eqref{eq:right_pm}, the GFT matrix with a first butterfly stage of $p$ Haar units has the form of
\begin{equation}
\label{eq:left_gft}
  \Um=\Bm_{n,p}\cdot\diag(\Um^+,\Um^-),\quad \Um^+\in \mathbb{O}^{n-p},\; \Um^-\in \mathbb{O}^{p}.
\end{equation}
%From \eqref{eq:left_gft} and the eigendecomposition $\Lm=\Um\Dm\Um^\top$ with a diagonal $\Dm$, 
Then the following lemma describes the conditions for $\Lm$ to have a GFT with $p$ left Haar units.

\begin{lemma}
\label{lem:graph_sym}
Let $\Lm$ be a graph Laplacian matrix, then there exists a GFT matrix $\Um$ in the form of \eqref{eq:left_gft}, i.e., the associated $\Gm_{YX}$, $\Gm_{YZ}$, $\Gm_{XY}$ and $\Gm_{ZY}$ are zero matrices, if and only if 
\begin{equation}
\label{eq:sym_condition}
  \Lm_{YY}=\Jm\Lm_{XX}\Jm,\quad \Lm_{YX}=\Jm\Lm_{XY}\Jm, \quad \Lm_{ZY}=\Lm_{ZX}\Jm.
\end{equation}
Note that when $Z$ is empty, i.e., $p=n/2$, \eqref{eq:sym_condition} implies that $\Lm$ has to be centrosymmetric, as in Lemma~\ref{lem:centrosym}.
\end{lemma}
\noindent {\it Proof:} 
If $\Um$ is a GFT matrix satisfying \eqref{eq:left_gft}, we denote the subblocks of $\Um^+$ as $\Um_{XX}^+$, $\Um_{XZ}^+$, $\Um_{ZX}^+$, and $\Um_{ZZ}^+$, and rewrite \eqref{eq:left_gft} as
\begin{equation}
\label{eq:umat_blocks}
  \Um=\frac{1}{\sqrt{2}}
  \begin{pmatrix}
    \Id_p & \zerov & \Jm_p \\ \zerov & \sqrt{2}\Id_{n-2p} & \zerov \\
    \Jm_p & \zerov & -\Id_p
  \end{pmatrix}\begin{pmatrix} 
  \Um_{XX}^+ & \Um_{XZ}^+ & \zerov \\ 
  \Um_{ZX}^+ & \Um_{ZZ}^+ & \zerov \\
  \zerov & \zerov & \Um^-
  \end{pmatrix}
\end{equation}
Denote the matrix of eigenvalues of $\Lm$ as $\Lambdam=\diag(\Lambdam_X,\Lambdam_Z,\Lambdam_Y)$ with subblock sizes $p$, $n-2p$, and $p$, respectively. Then, we can express each subblock of $\Lm$ by expanding $\Lm=\Um\Lambdam\Um^\top$ with \eqref{eq:umat_blocks}, and we can trivially verify that \eqref{eq:sym_condition} holds.

To show the converse, we assume that \eqref{eq:sym_condition} holds. Expanding the right hand side of \eqref{eq:BLB}, we can express the subblocks of $\Gm$ in terms of those of $\Lm$. In particular, 
\begin{align*}
  \Gm_{XY}&= \frac{1}{2}\left(\Lm_{XX}\Jm+\Jm\Lm_{YX}\Jm-\Lm_{XY}-\Jm\Lm_{YY}\right), \\
  \Gm_{ZY}&=\frac{\sqrt{2}}{2}\left(\Lm_{ZX}\Jm-\Lm_{ZY}\right).
\end{align*}
With these expressions, \eqref{eq:sym_condition} implies that $\Gm_{XY}$, $\Gm_{ZY}$, and their transpose versions $\Gm_{YX}$, $\Gm_{YZ}$ are all zero. This means that $\Gm$ is block-diagonal, and thus has an eigendecomposition as 
\[
  \Gm=\diag(\Vm_1,\Vm_2)\cdot\diag(\lambda_1,\dots,\lambda_n)\cdot\diag(\Vm_1,\Vm_2)^\top,
\]
where $\Vm_1\in\mathbb{O}^{n-p}$ and $\Vm_2\in\mathbb{O}^p$. It follows that $\Bm_{n,p}\cdot\diag(\Vm_1,\Vm_2)$ is an eigenmatrix of $\Lm=\Bm_{n,p}\cdot\Gm\cdot\Bm_{n,p}^\top$ as in \eqref{eq:left_gft}.
\qed

Under the conditions of \eqref{eq:sym_condition}, $\Gm$ reduces to 
\begin{equation}
\label{eq:blkdiag_G}
  \Gm =
  %\Bm_{n,p}^\top\Lm\Bm_{n,p}=
%  \begin{pmatrix}
%    \Lm_{XX}+\Lm_{XY}\Jm & \sqrt{2}\Lm_{XZ} & \zerov \\
%    \sqrt{2}\Lm_{XZ}^\top & \Lm_{ZZ} & \zerov \\
%    \zerov & \zerov & \Jm\Lm_{XX}\Jm - \Jm\Lm_{XY}.
%  \end{pmatrix}^\top=
  \begin{pmatrix}
    \Lm_{XX}+\Lm_{XY}\Jm & \sqrt{2}\Lm_{XZ} & \zerov \\
    \sqrt{2}\Lm_{XZ}^\top & \Lm_{ZZ} & \zerov \\
    \zerov & \zerov & \Lm_{YY} - \Jm\Lm_{XY}
  \end{pmatrix},
\end{equation}
and $\Um^+$ and $\Um^-$ are respectively the eigenmatrices of
\begin{equation}
\label{eq:Lpm}
  \Lm^+ := \begin{pmatrix}
    \Lm_{XX}+\Lm_{XY}\Jm & \sqrt{2}\Lm_{XZ} \\
    \sqrt{2}\Lm_{XZ}^\top & \Lm_{ZZ}
  \end{pmatrix}, \quad \Lm^- := \Lm_{YY} - \Jm\Lm_{XY}.
\end{equation}
A diagram with $n=8$, $p=3$ is shown in Fig.~\ref{fig:diagram}(b) as an example.

\begin{figure}[t]
\centering
\subfigure[$\Gc_1$]{
\includegraphics[width=.2\textwidth]{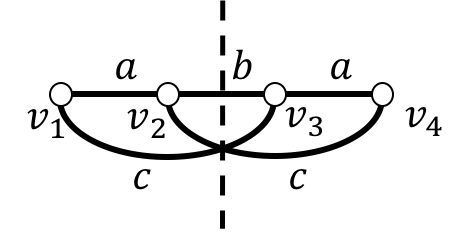}}%
\subfigure[$\Gc_2$]{
\includegraphics[width=.14\textwidth]{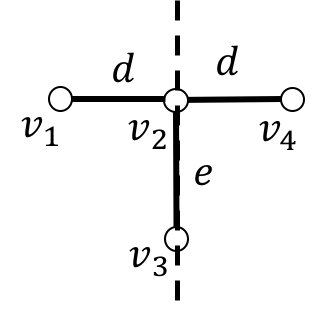}}
\caption{Example of symmetric graphs. (a) A graph with a bisymmetric Laplacian matrix. (b) A graph with a Laplacian satisfying \eqref{eq:sym_condition}.} 
\label{fig:sym_graph_example}
\end{figure}

Note that the desired properties in $\Lm$ correspond to certain symmetry properties \emph{in the graph topology}. If $\Vc_Z$ is empty, Lemma \ref{lem:graph_sym} implies that 
\begin{equation}
\label{eq:sym_L12}
w_{i,j}=w_{n+1-i,n+1-j},\quad \forall i\in\Vc,\; j\in\Vc.
\end{equation}
In this case, when we plot the nodes in order on a 1D line, we can identify an axis in the middle, around which all edges and self-loops are symmetric. An example of a graph whose Laplacian is bisymmetric is shown in Fig.~\ref{fig:sym_graph_example}(a). 

More generally, if $\Vc_Z$ is nonempty, then the first two equations in \eqref{eq:sym_condition} indicate that the sub-matrix of $\Lm$ associated to $\Vc_X$ and $\Vc_Y$ is bisymmetric. This means that $\Vc_X$ and $\Vc_Y$ contain vertices that are symmetric to each other. The third equation in \eqref{eq:sym_condition} implies that when there is an edge connecting $k\in\Vc_Z$ and $i\in\Vc_X$, there must be an edge with the same weight connecting $k$ and $n+1-i\in\Vc_Y$ as well. An example of this type of graph is shown in Fig.~\ref{fig:sym_graph_example}(b), where $\Vc_X=\{1\}$, $\Vc_Z=\{2,3\}$, and $\Vc_Y=\{4\}$. Similar to Fig.~\ref{fig:sym_graph_example}(a), we can identify a symmetry around the middle, though nodes in $\Vc_Z$ are not paired with symmetric counterparts. 

Based on the observations above, we see that left Haar units are available when the graph has symmetry properties related to Lemma \ref{lem:graph_sym}. However, Lemma \ref{lem:graph_sym} assumes that the nodes had been ordered properly, so that the Laplacian has the required bisymmetric structure. In general, the node labels of a graph will not be such that this condition is automatically met, even if the graph is symmetric. For example, if a different node labeling is applied to the graph of Fig. \ref{fig:sym_graph_example}(a), its corresponding Laplacian may not be bisymmetric anymore. In the next section we study methods to identify graph symmetries directly using node pairing functions, which will allow us to design fast GFT algorithms, regardless of how the nodes are initially labeled.

\section{Fast GFTs Based on Graph Symmetry}
\label{sec:sym_graph}
In this section, we will characterize how Lemma~\ref{lem:graph_sym} relates to the graph topology. In particular, we define the symmetry properties observed in Fig.~\ref{fig:sym_graph_example} based on an involution (node-pairing function) in Section \ref{subsec:involution}. Given an observed graph symmetry characterized by an involution, in Section \ref{subsec:node_partition} we define the node sets $\Vc_X$, $\Vc_Y$, and $\Vc_Z$. In Section \ref{subsec:decomposition}, we propose a graph decomposition approach for searching fast GFTs in stages. %Finally, in Section \ref{subsec:interpretation}, we provide some interpretations of the graphs obtained in the proposed approach.

\subsection{Graph Symmetry Based on Node Pairing}
\label{subsec:involution}
The symmetries of Fig.~\ref{fig:sym_graph_example} can be described in terms of complete  (Fig.~\ref{fig:sym_graph_example}(a)) and incomplete (Fig.~\ref{fig:sym_graph_example}(b)) {\em node pairings}. Such pairings can be defined by bijective mappings that are their own inverses, namely, \emph{involutions}:
%\begin{definition}[\cite{pemmaraju2003computational}]
\begin{definition}[\cite{knuth1998art}]
A permutation on a finite set $\Vc$ is called an involution if it is its own inverse, i.e., $\phi(\phi(i))=i$ for all $i\in\Vc$. 
\end{definition}
We will use them to identify graph symmetries. 
\begin{definition}
\label{def:graph_sym}
Let $\phi$ be an involution on the vertex set $\Vc$ of a graph $\Gc$, then $\Gc$ is $\phi$-symmetric if $w_{i,j}=w_{\phi(i),\phi(j)}$ for all $i\in\Vc$, $j\in\Vc$.
\end{definition}
Note that in Definition \ref{def:graph_sym}, the required property has to hold also for $i=j$. That is, $s_i=w_{i,i}=w_{\phi(i),\phi(i)}=s_{\phi(i)}$, meaning that the self-loops on nodes $i$ and $\phi(i)$ are required to have the same weight. Also note that, among permutations, only involutions are valid for Definition \ref{def:graph_sym}, since the pairing functions that lead to the conditions in \eqref{eq:sym_condition} can only be induced by involutions.\footnote{Note that graph symmetry can be defined differently in different contexts. In algebraic graph theory, graph symmetry is defined based on transitivity of vertices and edges \cite{godsil2001algebraic}. In \cite{erdos1963asymmetric}, a graph is called symmetric if there exists a non-identical permutation $\phi$ (not necessarily an involution) on the graph nodes that leaves the graph unaltered. These definitions are beyond the scope of this paper. When we refer to graph symmetry in this paper, we always assume an involution $\phi$ is specified such that Definition \ref{def:graph_sym} holds.}
%Such symmetry properties are investigated in order to explore divide and conquer implementations for the GFT.

Let $\Vc=\{1,\dots,n\}$ be the vertex set, and let us denote an involution $\phi$ as $\phi=(\phi(1),\phi(2),\dots,\phi(n))$. For example, the involutions corresponding to the symmetries of graphs in Figs.~\ref{fig:sym_graph_example}(a) and (b) are $\phi_a=(4,3,2,1)$ and $\phi_b=(4,2,3,1)$, respectively. We also denote the number of available Haar units for a given $\phi$ as 
\begin{equation}
\label{eq:p_phi}
  p_\phi := \frac{1}{2}\times\left|\{i\in\Vc:\; i\neq\phi(i)\}\right|.
\end{equation}

\subsection{Node Partitioning for Haar Units}
\label{subsec:node_partition}
Once we observe a graph symmetry and characterize it by an involution $\phi$, we can identify the nodes on the axis of symmetry, $\Vc_Z:=\{i\in\Vc:\; \phi(i)=i\}$, then partition the other nodes into two sets $\Vc_X$ and $\Vc_Y$ such that nodes in those sets belong to different sides of the symmetry axis. In this way, we can define an orthogonal matrix $\Bm_{\phi}$ as a permuted version of $\Bm_{n,p_\phi}$ based on $\phi$, $\Vc_X$, $\Vc_Y$, and $\Vc_Z$ in the following way:
\begin{equation}
\label{eq:B_phi}
  (\Bm_{\phi})_{i,j}=\left\{\begin{array}{ll}
  1/\sqrt{2}, & i=j\in\Vc_X \\
  -1/\sqrt{2}, & i=j\in\Vc_Y \\
  1, & i=j\in\Vc_Z \\
  1/\sqrt{2}, & i\in\Vc_X, \; j=\phi(i)\in\Vc_Y\\
  1/\sqrt{2}, & i\in\Vc_Y, \; j=\phi(i)\in\Vc_X\\
  0, & \text{otherwise}
  \end{array}\right.
\end{equation}
This means that $\Lm_{\phi} := \Bm_{\phi}^\top\Lm\Bm_{\phi}$ is a permuted version of \eqref{eq:blkdiag_G}, whose block diagonal structure gives the following theorem:
\begin{theorem}[Block-diagonalization of Laplacian based on graph symmetry]
\label{thm:blockdiag}
Let the graph $\Gc$ with Laplacian $\Lm$ be $\phi$-symmetric. Then, $(\Lm_{\phi})_{i,j}=(\Lm_{\phi})_{j,i}=0$ if $i\in\Vc_X\cup\Vc_Z$ and $j\in\Vc_Y$.
\end{theorem}
While Theorem \ref{thm:blockdiag} is derived based on the unnormalized Laplacian $\Lm$, it holds for normalized Laplacian as well.

\subsection{Main Approach--Decomposition of Symmetric Graphs}
\label{subsec:decomposition}

The block-diagonalization of \eqref{eq:BLB} maps $\Lm$ to $\Gm$ via $\Bm_{n,p}$, with $\Gm$ in \eqref{eq:blkdiag_G}. Note that, from \eqref{eq:sw_from_l} and \eqref{eq:l_from_sw}, we can draw a one-to-one correspondence between a matrix and a graph. In this way, $\Gm=\diag(\Lm^+,\Lm^-)$ can be can regarded as the Laplacian of a graph with two connected components, denoted as $\Gc^+$ and $\Gc^-$, with Laplacians $\Lm^+$ and $\Lm^-$, vertex sets $\Vc^+ := \Vc_X\cup\Vc_Z$ and $\Vc^- := \Vc_Y$, weight matrices $\Wm^+$ and $\Wm^-$ (possibly with negative weights), respectively. With this graph decomposition from $\Gc$ to $\Gc^+$ and $\Gc^-$, the GFT of $\Gc$ can be implemented by a butterfly stage $\Bm_{n,p}$, followed by the two \emph{sub-GFTs} corresponding to $\Gc^+$ and $\Gc^-$. Explicitly considering the graphs resulting from this decomposition is useful because in some cases $\Lm^+$ and $\Lm^-$ may in turn have symmetry properties, which could be  exploited to achieve additional reductions in complexity. Moreover, considering the transforms after the Haar units as GFTs could lead to better interpretations of the overall GFT. 

% By definition of graph Laplacian, we can express the self-loop weights $s_i$ and edge weights $w_{i,j}$ in terms of entries of the Laplacian matrix $\Lm=(l_{i,j})_{i,j}$, and vice versa:
% \begin{align}
% \label{eq:sw_from_l}
%   & s_i=\sum_{j\in\Vc}l_{i,j},\quad w_{i,j}=-l_{i,j}\text{ for } i\neq j, \\
% \label{eq:l_from_sw}
%   & l_{ii} = s_i + \sum_{\substack{j\in\Vc \\ j\neq i}} w_{i,j},\quad l_{i,j}=-w_{i,j}\text{ for } i\neq j.
% \end{align}
Regarding $\Lm^+$ and $\Lm^-$ in \eqref{eq:Lpm} as graph Laplacians, we can use \eqref{eq:sw_from_l} and \eqref{eq:l_from_sw} to express self-loop and edge weights of $\Gc^+$ and $\Gc^-$ in terms of those of $\Gc$, as described in the following theorem.
% \begin{lemma} If $\Gc$ is $\phi$-symmetric with node partitions $\Vc_X=\Vc^+$, $\Vc_Y=\Vc^-$, and $\Vc_Z=\emptyset$, then
% \label{lem:d1}
% \begin{align*}
% %\label{eq:wp1}
%   & w_{i,j}^+ = w_{i,j}+w_{i,\phi(j)}, \quad \forall i,j\in\Vc^+,\quad i\neq j, \\
% %\label{eq:sp1}
%   & s_i^+ = s_i, \quad \forall i\in\Vc^+ \\
% %\label{eq:wm1}
%   & w_{i,j}^- = w_{i,j}-w_{i,\phi(j)}, \quad \forall i,j\in\Vc^-,\quad i\neq j, \\
% %\label{eq:sm1}
%   & s_i^- = s_i + 2\sum_{j\in\Vc^+} w_{i,j}, \quad \forall i\in\Vc^-.
% \end{align*}
% \end{lemma}
\begin{theorem} If $\Gc$ is $\phi$-symmetric with node partitions $\Vc_X$, $\Vc_Y$, and $\Vc_Z$, then the weights of $\Gc^+$ (with vertex set $\Vc^+=\Vc_X\cup\Vc_Z$) and $\Gc^-$ (with vertex set $\Vc^-=\Vc_Y$) are given by
\label{thm:decompose}
\begin{align*}
%\label{eq:wp2}
  & w_{i,j}^+ = \left\{\begin{array}{ll} 
    w_{i,j}+w_{i,\phi(j)}, & \text{ if } i\in\Vc_X,\; j\in\Vc_X \\
    \sqrt{2}w_{i,j}, & \text{ if } i\in\Vc_X, \; j\in\Vc_Z \text{ or } i\in\Vc_Z, \; j\in\Vc_X \\
    w_{i,j}, & \text{ if } i\in\Vc_Z, \; j\in\Vc_Z,
  \end{array}\right. \\
%\label{eq:sp2}
  & s_i^+ = \left\{\begin{array}{ll} 
    s_i-(\sqrt{2}-1)\sum_{j\in\Vc_Z}w_{i,j}, & \text{ if }i\in\Vc_X \\
    s_i+(2-\sqrt{2})\sum_{j\in\Vc_X}w_{i,j}, & \text{ if }i\in\Vc_Z, 
  \end{array}\right. \\
%\label{eq:wm2}
  & w_{i,j}^- = w_{i,j}-w_{i,\phi(j)},\quad \forall i,j\in\Vc_Y,\quad i\neq j \\
%\label{eq:sm2}
  & s_i^- = s_i + 2\sum_{j\in\Vc_X} w_{i,j} + \sum_{j\in\Vc_Z} w_{i,j},\quad \forall i\in\Vc_Y.
\end{align*}
\end{theorem}
%Note that, by symmetry, we also have $w_{i,j}^+= w_{i,j}+w_{\phi(i),j}$ for \eqref{eq:wp1} and \eqref{eq:wp2}, and $w_{i,j}^-=w_{i,j}-w_{\phi(i),j}$ for \eqref{eq:wm1} and \eqref{eq:wm2}. 
%The proof of Lemma \ref{thm:decompose}, the more general version of Lemma \ref{lem:d1}, refers to Appendix \ref{app:d2}.
%The proof of this theorem is in Appendix \ref{app:d2}. Based on the results above, $\Gc^+$ and $\Gc^-$ can be obtained from a symmetric graph $\Gc$ as follows. 

% \begin{figure}[t]
% \centering
% \subfigure[$\Gc$]{
% \includegraphics[scale=.4]{figures/Gl.png}}
% \subfigure[$\Gc^+$]{
% \includegraphics[scale=.4]{figures/Gl_p.png}}
% \subfigure[$\Gc^-$]{
% \includegraphics[scale=.4]{figures/Gl_m.png}}
% \caption{Symmetric graph decomposition for the graph in Fig.~\ref{fig:sym_graph_example}(a). Red diamonds and blue squares represent nodes in $\Vc_X$ and $\Vc_Y$, respectively. Note that the modifications based on weights $a$ and $b$ are different, as one connects symmetric nodes, and the other does not.} 
% \label{fig:decompose42}
% \end{figure}

% \begin{figure}
% \centering
% \subfigure[$\Gc$]{
% \includegraphics[scale=.4]{figures/Gm.png}}%
% \subfigure[$\Gc^+$]{
% \includegraphics[scale=.4]{figures/Gm_p.png}}%
% \subfigure[$\Gc^-$]{
% \includegraphics[scale=.4]{figures/Gm_m.png}}
% \caption{Symmetric graph decomposition for the graph in Fig.~\ref{fig:sym_graph_example}(b). Red diamonds, green triangles, and blue squares represent nodes in $\Vc_X$, $\Vc_Z$, and $\Vc_Y$, respectively.}
% \label{fig:decompose41}
% \end{figure}

\begin{figure}[t]
\centering
\subfigure[]{
\includegraphics[scale=.36]{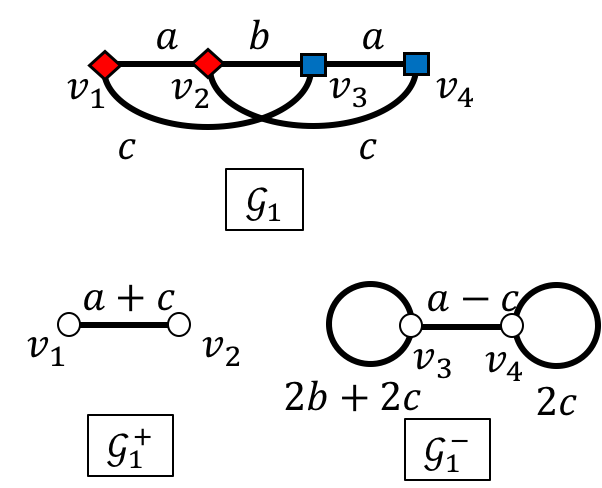}}\hspace{.1cm}
\subfigure[]{
\includegraphics[scale=.4]{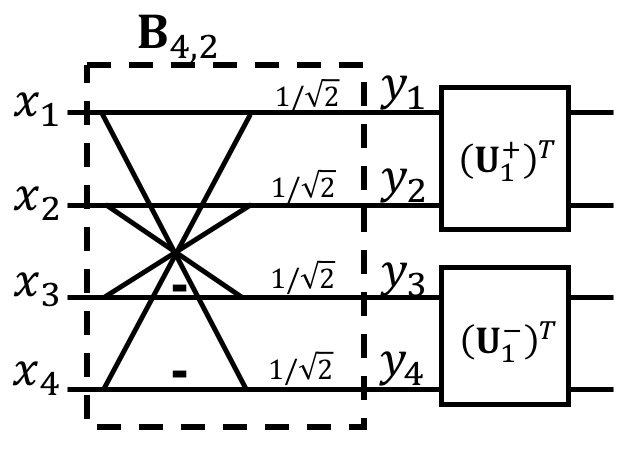}}
\caption{(a) Symmetric graph decomposition for the graph in Fig.~\ref{fig:sym_graph_example}(a). Red diamonds and blue squares represent nodes in $\Vc_X$ and $\Vc_Y$, respectively. (b) The associated fast GFT diagram for $\Gc_1$, where $\Um_1^+$ and $\Um_1^-$ are the GFTs of $\Gc_1^+$ and $\Gc_1^-$, respectively.} 
\label{fig:decompose42}
\end{figure}

\begin{figure}
\centering
\subfigure[]{
\includegraphics[scale=.36]{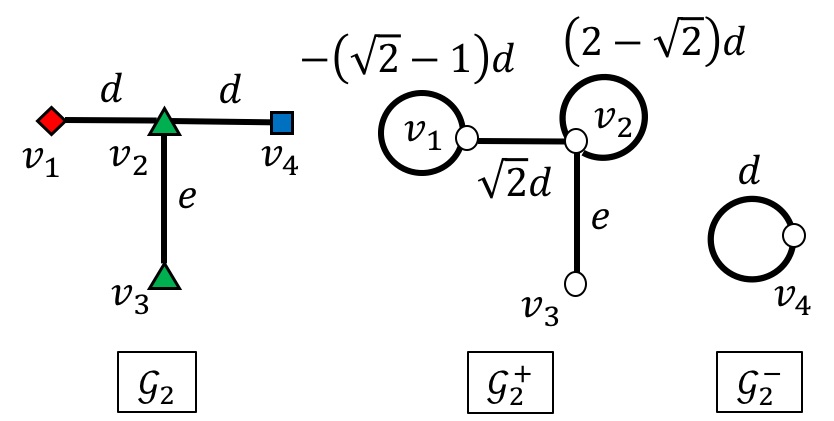}}%
\subfigure[]{
\includegraphics[scale=.36]{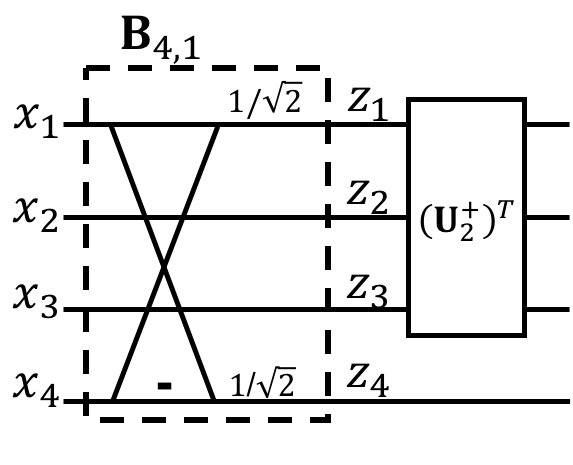}}
\caption{(a) Symmetric graph decomposition for the graph in Fig.~\ref{fig:sym_graph_example}(b). Red diamonds, green triangles, and blue squares represent nodes in $\Vc_X$, $\Vc_Z$, and $\Vc_Y$, respectively. (b) The associated fast GFT diagram for $\Gc_2$, where $\Um_2^+$ is the GFT of $\Gc_2^+$.}
\label{fig:decompose41}
\end{figure}

Refer to Appendix \ref{app:d2} for the proof. 
% \subsection{Interpretations}
% \label{subsec:interpretation}
We use the toy examples of Figs.~\ref{fig:decompose42} and \ref{fig:decompose41} (with $\Vc_Z=\emptyset$ and $\Vc_Z\neq\emptyset$, respectively)  to illustrate the graph decomposition. Note that $\Gc^+$ and $\Gc^-$ may have negative weights even if $\Gc$ does not. For any signal $\xv$, we denote the ``sum'' (low-pass) and ``difference'' (high-pass) outputs of Haar units as $\xv^+$ and $\xv^-$: $((\xv^+)^\top,(\xv^-)^\top)^\top=\Bm_\phi^\top\xv$. For example, in Fig.~\ref{fig:decompose42}, $\xv^+=(y_1,y_2)^\top$ and $\xv^-=(y_3,y_4)^\top$. In Fig.~\ref{fig:decompose41}, $\xv^+=(z_1,z_2,z_3)^\top$ and $\xv^-=(z_4)$. 

The graph construction of Theorem~\ref{thm:decompose} creates two disconnected sub-graphs by removing all edges between $\Vc_y$ and $\Vc_x\cup\Vc_z$ and preserving all other edges, but changing some of the weights and adding self-loops. Three types of cases lead to one or two edges being removed:
% Thus, 
% \[
%   \xv^\top\Lm\xv = \xv^\top 
%   \left[ \Bm_\phi \begin{pmatrix} \Lm^+ & \zerov \\ \zerov & \Lm^- \end{pmatrix} \Bm_\phi^\top \right] \xv 
%   = (\yv^+)^\top\Lm^+\yv^+ + (\yv^-)\top\Lm^-\yv^-,
% \]
% where $\yv^+\Lm^+\yv^+$ and $\yv^-\Lm^-\yv^-$ are variations of signals $\yv^+$ and $\yv^-$ on $\Gc^+$ and $\Gc^-$, respectively. This means that the decomposition of $\Gc$ into $\Gc^+$ and $\Gc^-$ also divides the graph variation of $\xv$ on $\Gc$ into variations on $\Gc^+$ and $\Gc^-$, with corresponding sub-GFTs $\Um^+$ and $\Um^-$.

\begin{figure}
    \centering
    \includegraphics[width=.46\textwidth]{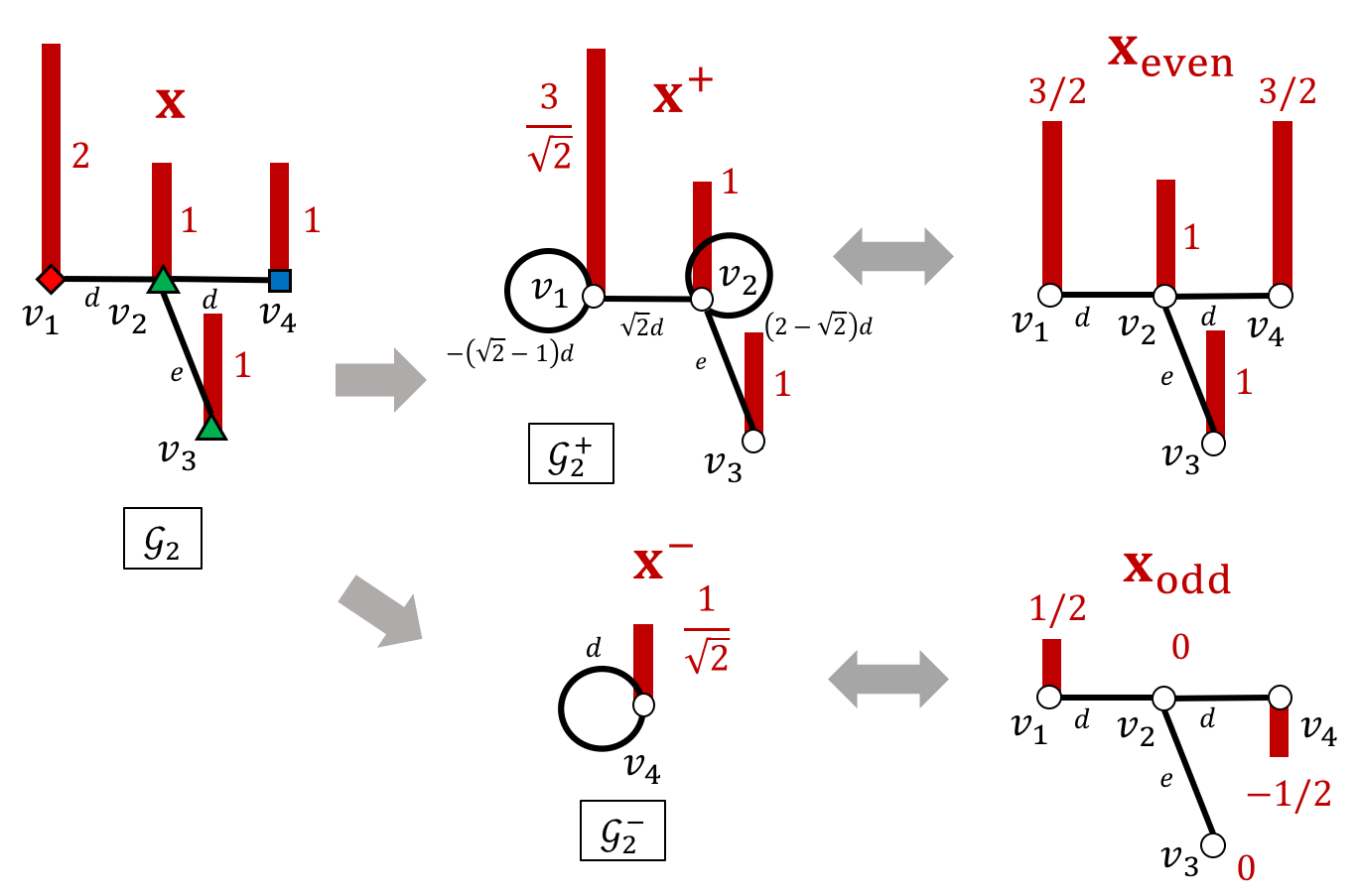}
    \caption{An example of even and odd symmetric components on the graph in Fig.~\ref{fig:decompose41}. Signals $\xv^+$ and $\xv^-$ are outputs of the Haar units. Signals $\xv_\text{even}$ and $\xv_\text{odd}$ are associated to $\xv^+$ and $\xv^-$ by \eqref{eq:even_odd}.}
    \label{fig:x_even_odd_41}
\end{figure}

\begin{enumerate}
    \item {\bf Edges connecting two symmetric nodes $i\in\Vc_X$ and $\phi(i)\in\Vc_Y$}. The edge with weight $b$ in Fig. \ref{fig:decompose42}(a) is an example of this case. These edges are removed and lead to   self-loops with twice the original weight in $\Gc^-$ ($2b$ in this case).
    
    \item {\bf Two symmetric edges: each connecting a node in $\Vc_X$ to a node in $\Vc_Y$}. The two edges with weight $c$ in Fig. \ref{fig:decompose42}(a) are an example. These edges are removed, but lead to changes in two edge weights, with the weight of the edge in $\Gc^+$ increasing and that of the edge in $\Gc^-$ decreasing. Two self-loops are also added to the corresponding nodes in $\Gc^-$.
    
    \item {\bf Two symmetric edges with a common node in $\Vc_Z$}. Edges with weight $d$ in Fig. \ref{fig:decompose41}(a) belong to this case. This case results in a single edge being kept, with a modified edge weight and two self-loops in $\Gc^+$, and a self-loop in $\Gc^-$.
\end{enumerate}

Note that, the signals $\xv^+$ and $\xv^-$ correspond to $\Gc^+$ and $\Gc^-$, and can be regarded as even and odd symmetric components of the original graph signal $\xv$. An example associated to Fig.~\ref{fig:decompose41} is shown in Fig.~\ref{fig:x_even_odd_41}. We can see that a graph signal can be decomposed into two components $\xv_\text{even}$ and $\xv_\text{odd}$, which correspond to $\xv^+$ and $\xv^-$, respectively, by
\begin{equation}\label{eq:even_odd}
\scriptstyle
  \xv_\text{even}(i) = \left\{\begin{array}{ll}
       \scriptstyle \xv^+(i)/\sqrt{2}, & \scriptstyle i\in\Vc_X \\
       \scriptstyle \xv^+(\phi(i))/\sqrt{2}, & \scriptstyle i\in\Vc_Y \\
       \scriptstyle \xv^+(i), & \scriptstyle i\in\Vc_Z
  \end{array}\right. ,\quad
  \xv_\text{odd}(i) = \left\{\begin{array}{ll}
       \scriptstyle \xv^-(\phi(i))/\sqrt{2}, & \scriptstyle i\in\Vc_X \\
       \scriptstyle -\xv^-(i)/\sqrt{2}, & \scriptstyle i\in\Vc_Y \\
       \scriptstyle 0, & \scriptstyle i\in\Vc_Z
  \end{array}\right.
\end{equation}
In particular, $\xv_\text{even}$ and $\xv_\text{odd}$ have even and odd symmetries based on the node pairing, i.e. $\xv_\text{even}(i)=\xv_\text{even}(\phi(i))$ and $\xv_\text{odd}(i)=-\xv_\text{odd}(\phi(i))$ for all $i\in\Vc$. This can be considered as a generalization of even and odd symmetric components decomposition for finite length time series. Components $\xv_\text{even}$ and $\xv_\text{odd}$ of the graph signal $\xv$ can be regarded as intermediate results of the GFT coefficients.

The decomposition described in Theorem \ref{thm:decompose} enables us to search further stages of Haar units in the sub-GFTs $\Um^+$ and $\Um^-$ by inspecting their associated graphs $\Gc^+$ and $\Gc^-$. Once a symmetry based on an involution is found in $\Gc^+$ or $\Gc^-$, we can apply the decomposition again, and repeat until a symmetry property cannot be found anymore. Some examples will be provided in Section \ref{sec:examples}. %Additional stages of Haar units are more likely to be available for graphs that have regular topologies and/or uniform weights. 

\section{Examples and Applications}
\label{sec:examples}
In practice, graphs with distinct weights on different edges or graphs learned from data without any topology constraints are not likely to have the desired bipartition and symmetry properties. 
%Searching for an involution that induces symmetry in a given graph may be a combinatorial problem since the number of involutions grows with $n$ at a rate faster than a polynomial in $n$ \cite{pemmaraju2003computational}. 
However, bipartite and symmetric graph structures arise in graphs considered in certain fields. Examples of bipartite graphs include tree-structured graphs, whose GFTs are useful for designing wavelet transforms on graph \cite{shen2009tree-based}. Involution-based symmetries can be found in graphs with regular or partially regular topologies (e.g. line, cycle, and grid graphs), graphs that are symmetric by construction (e.g. human skeletal graphs), and uniformly weighted graphs. In what follows, %Secs.~\ref{subsec:example_bipartite}-\ref{subsec:example_skeletal}, 
we study several classes of graphs with these properties and discuss the search of involution in general graphs.

% \subsection{Bipartite Graphs}
% \label{subsec:example_bipartite}
% As stated in Lemma \ref{lem:k-rbg}, a k-regular bipartite graph has a GFT with a right butterfly stage. Let the sizes of the parts in the bipartite graph be $|\Sc_1|=p$, $|\Sc_2|=n-p$, then sub-GFTs $\Em$ and $\Fm$ have sizes $p$ and $n-p$, and the total number of multiplication operations will be $p^2+(n-p)^2$. By Lemma \ref{lem:norm-bg}, the same result can be derived for a GFT derived from the normalized Laplacian of any bipartite graph.

\subsection{Graphs with 2-Sparse Eigenvectors}
The authors in \cite{teke2017uncertainty} have studied the conditions for 2-sparse graph eigenvectors to exist:
\begin{lemma}[\cite{teke2017uncertainty}]
\label{lem:2sparse}
A Laplacian has an eigenvector $\uv$ with only two nonzero elements
$\uv(i)=1/\sqrt{2}$, $\uv(j)=-1/\sqrt{2}$ if and only if 
\begin{equation}
\label{eq:2sparse}
  \forall v\in\Vc \backslash \{i,j\}, \quad w_{v,i}=w_{v,j}.
\end{equation}
\end{lemma}
In fact, the condition \eqref{eq:2sparse} is equivalent to having $\Gc$ $\phi$-symmetric, with $\phi(i)=j$, $\phi(j)=i$, and $\phi(k)=k$ for $k\neq i,j$. In this case, each of $\Vc_X=\{i\}$ and $\Vc_Y=\{j=\phi(i)\}$ has only one node, and $\Um^-$ reduces to a one by one identity matrix. 
Examples of graphs satisfying Lemma \ref{lem:2sparse} include uniformly weighted graphs with several types of topology: 1) star graph, 2) complete graph, and 3) a graph with a clique (a complete subgraph), where at least two nodes in the clique are not connected to any other nodes outside the clique. 

\begin{figure*}[th]
\centering
\subfigure[]{
\includegraphics[width=.41\textwidth,valign=b]{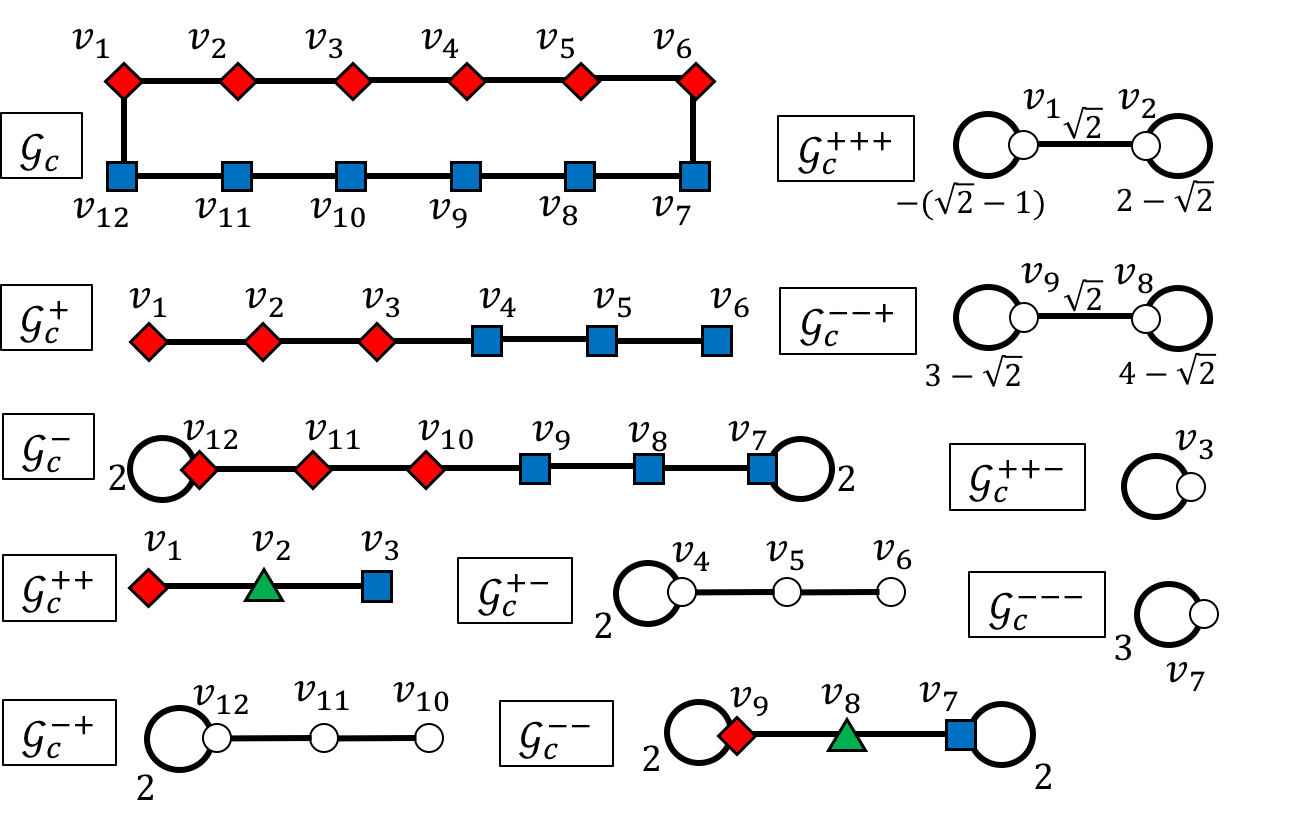}}%
\subfigure[]{
\includegraphics[width=.57\textwidth,valign=b]{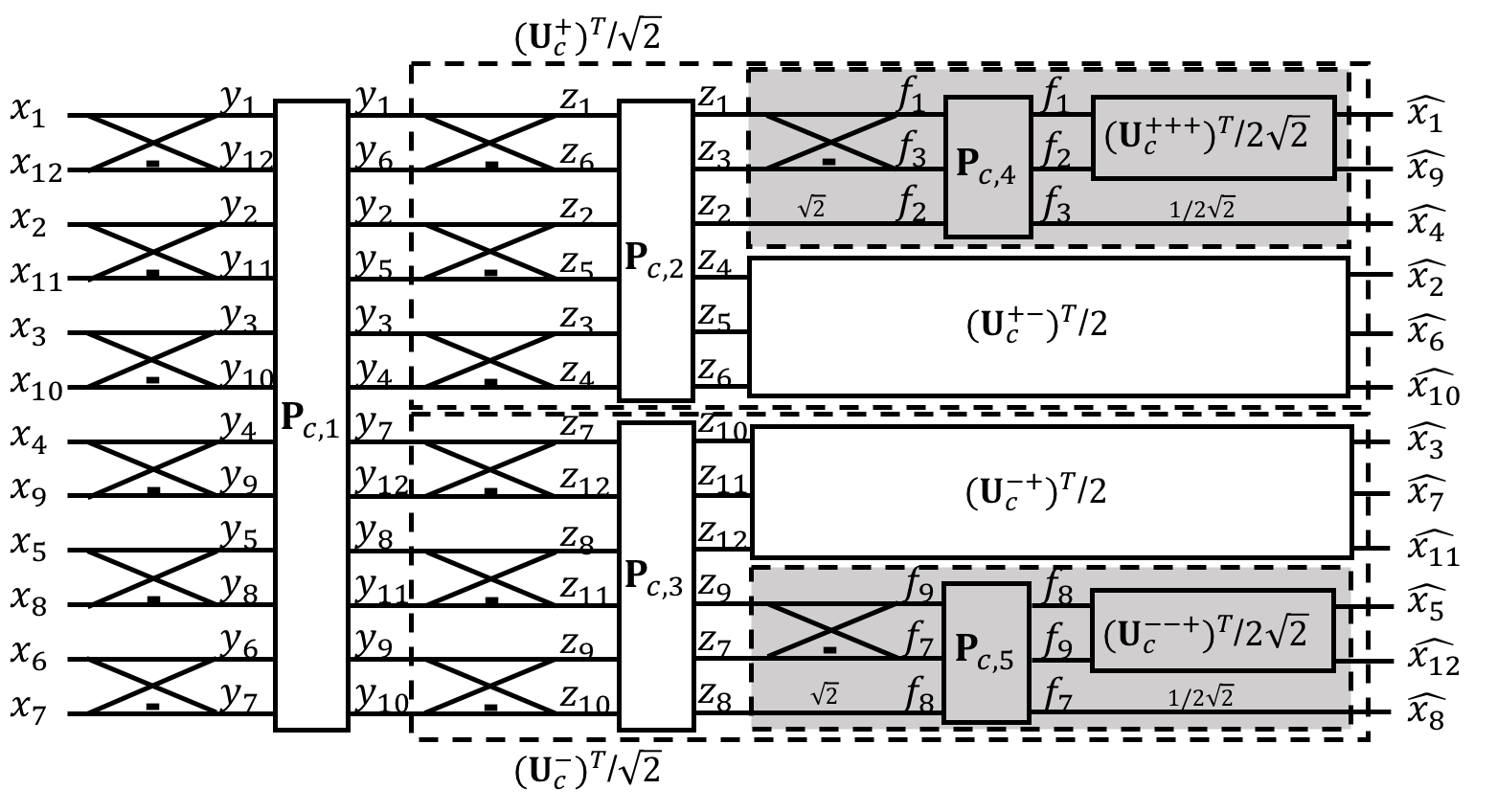}}
\caption{The 12-node cycle graph: (a) Graph decomposition. (b) The associated fast GFT diagram. Red diamonds, green triangles, and blue squares represent those nodes in $\Vc_X$, $\Vc_Z$, and $\Vc_Y$, respectively, for the \emph{next} stage of decomposition. Unlabeled edges and self-loops have weights 1, and $\Pm_{c,i}$ are permutation operations. The two shaded sub-GFTs are $(\Um_c^{++})^\top/2$ (top) and $(\Um_c^{--})^\top/2$ (bottom), respectively.}
\label{fig:cycle12}
\end{figure*}

\subsection{Symmetric Line Graphs}
A Laplacian matrix associated to a line graph can be viewed as a precision matrix (inverse covariance matrix) of a first-order Gaussian Markov random field (GMRF), which can be used for modeling image and video pixels \cite{li2009markov,zhang2015graph}. One special case is the line graph with uniform weights, whose GFT is the well-known DCT. 

If a line graph $\Gc_l$ is symmetric around the middle, then it is $\phi=(n,n-1,\dots,1)$-symmetric $\Gc_l$ and has a left butterfly stage. In our recent work \cite{lu2016symmetric}, we consider inter-predicted residual blocks in video coding, where pixels in a residual block have nearly symmetric statistics around the middle. We model those blocks by a GMRF based on a symmetric line graph. The resulting GFT has a fast implementation and provides a coding gain as compared to the DCT.

\begin{figure*}[th]
%\vspace{-.5cm}
\centering
\subfigure[]{
\includegraphics[width=.48\textwidth,valign=b]{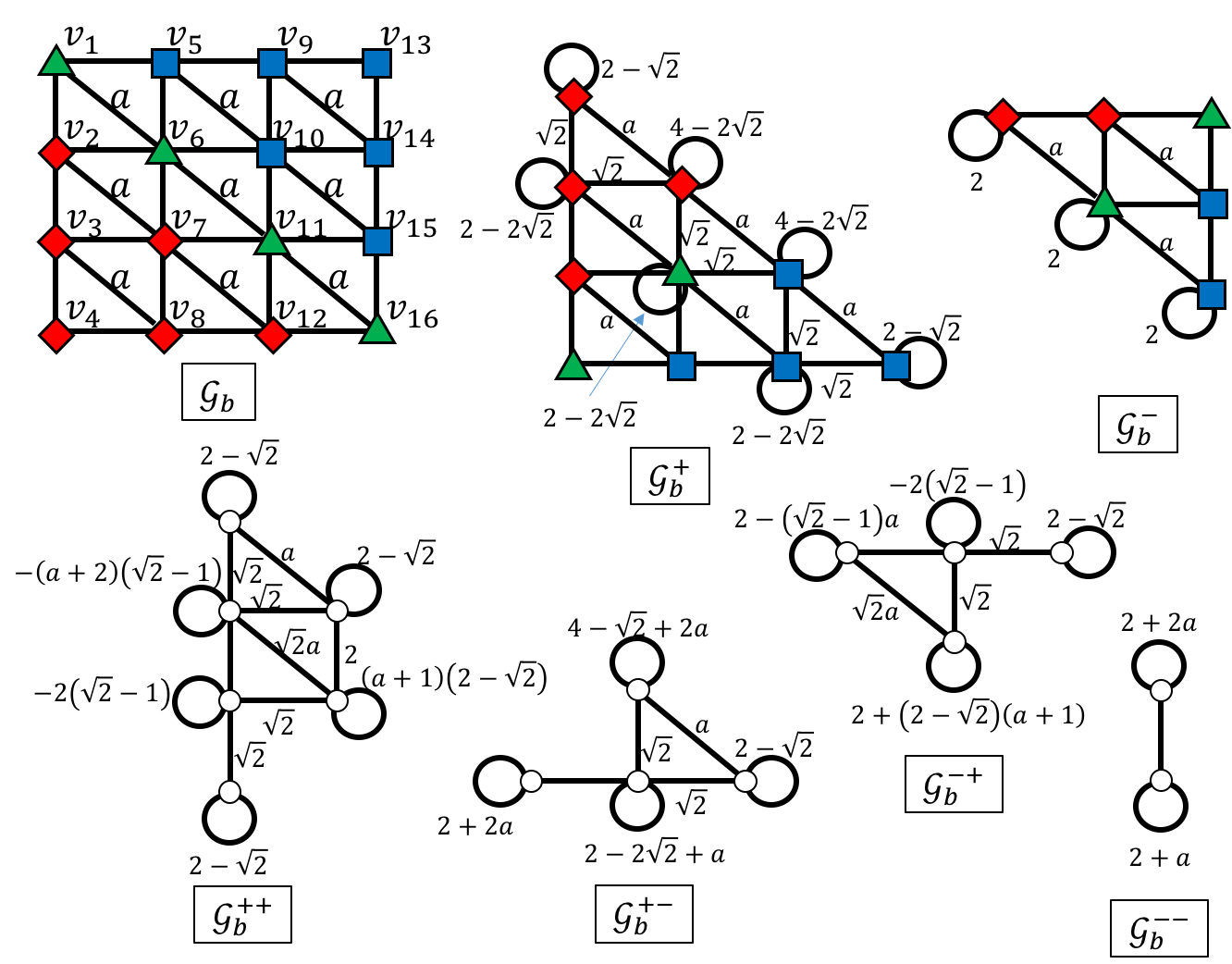}}
\hspace{.5cm}
\subfigure[]{
\includegraphics[width=.46\textwidth,valign=b]{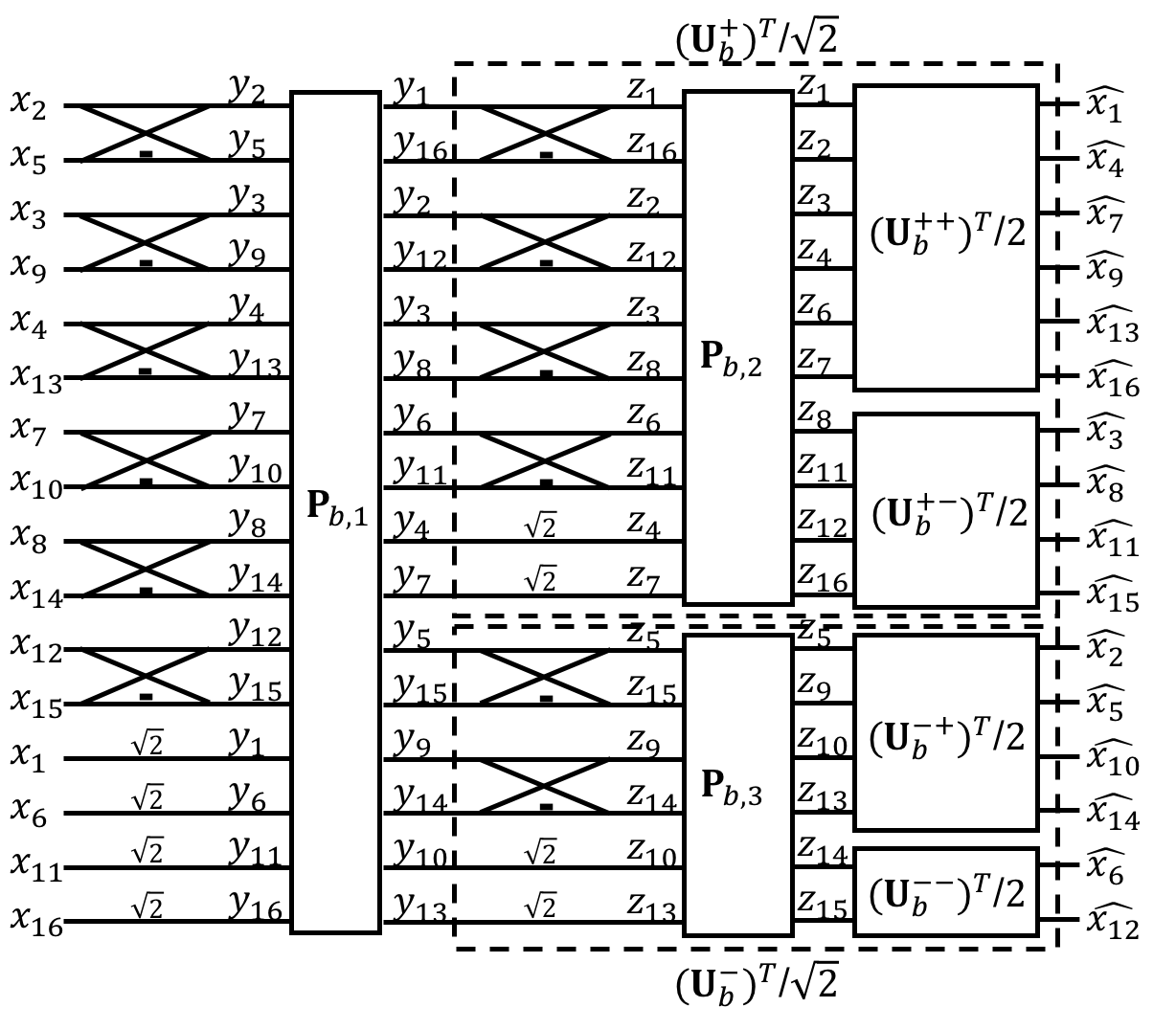}}
\caption{The bi-diagonally symmetric grid. (a) Graph decomposition. (b) The associated fast GFT diagram. Red diamonds, green triangles, and blue squares represent those nodes in $\Vc_X$, $\Vc_Z$, and $\Vc_Y$, respectively, for the \emph{next} stage of decomposition. Unlabeled edges have weights 1, and $\Pm_{b,i}$ are permutation operations.}
\label{fig:grid_bidiag}
\end{figure*}

\subsection{Steerable DFTs}
The Laplacian of an $n$-node cycle graph $\Gc_c$ with unit weights is circulant and has non-unique GFTs since some of its eigenvalues have multiplicities greater than one. Due to the circulant structure, the DFT is one of the GFTs of $\Gc_c$. The family of all GFTs of $\Gc_c$ is called the steerable DFTs \cite{fracastoro2017steerable_dft}.
Note that, for any $n$, $\Gc$ is $\phi=(n,n-1,\dots,3,2,1)$-symmetric, which enables us to explore fast implementations for $\Gc_c$ other than the Cooley-Tukey fast Fourier transform (FFT) algorithm \cite{cooley1965algorithm}. 
%In fact, the transform shown in Fig.~\ref{fig:diagram_c4}(b) can be obtained using the proposed graph decomposition. 
An example with $n=12$ is shown in Fig. \ref{fig:cycle12}, where two stages of Haar units are available, and some of the sub-GFTs after the first two stages can be further simplified. Unlike the conventional DFT, the derived GFT will have real operations only. In addition, while FFT cannot be easily applied when $n$ is a prime number, our method gives at least one left butterfly stage for any $n$. We also note that, for any steerable DFT with a length $n$ that is a multiple of 4, the GFTs of $\Gc_c^{++}$ and $\Gc_c^{-+}$ are Type-2 DCT and Type-4 DST, respectively. This means that those sub-GFTs can also be implemented using fast DCT and ADST algorithms \cite{feig1992fast,kok1997fast,han2013butterfly}. To the best of our knowledge, fast implementations for steerable DFT other than the FFT algorithm have not been studied in the literature.

\subsection{Symmetric Grid Graphs}

\begin{figure}[t]
    \centering
    \subfigure[]{
    \includegraphics[height=.3\textwidth]{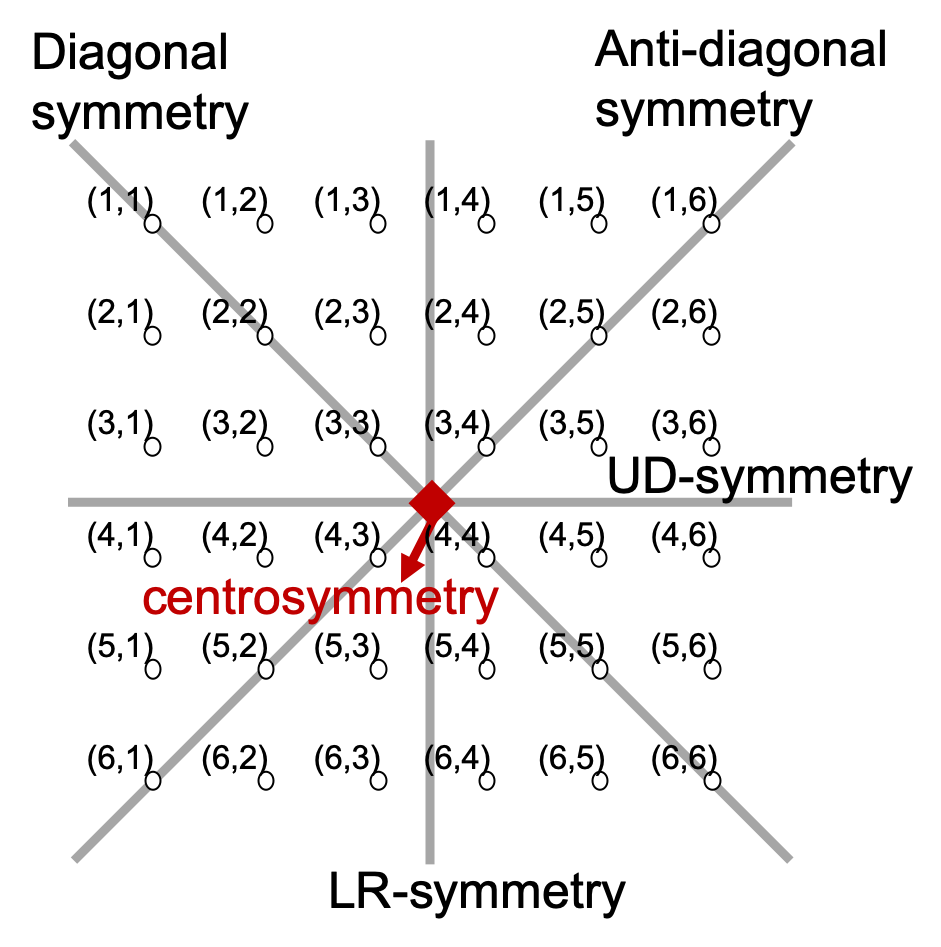}}%
    \subfigure[]{
    \includegraphics[height=.3\textwidth]{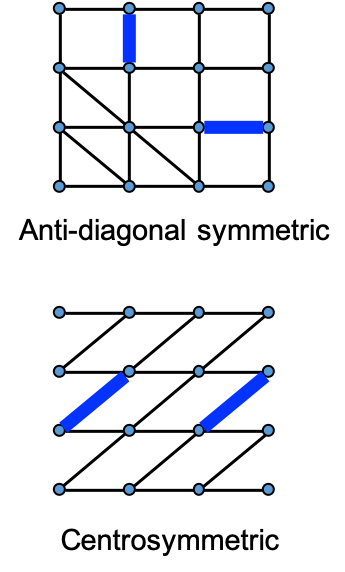}}
    \caption{(a) Axes and point of symmetry for different symmetry types of a 6$\times$6 grid. Node indices are represented based on the image coordinate system. (b) Examples of anti-diagonal and centrosymmetric grids, where the bold colored edges are symmetric to each other based on axis/point of symmetry depicted in (a).}
    \label{fig:grid_coords}
\end{figure}

For the term ``grid'', we refer to a graph with $n=N_1\times N_2$ nodes that correspond to integer positions in 2D Euclidean space. This means that each node can be associated to a 2D coordinate $(k,l)$ with $1\leq k\leq N_1$ and $1\leq l \leq N_2$. In practice, many grids that arise in applications such as image processing have highly regular topologies due to the structured data domain. For example, pixel data can be modeled by a 4-connected grid, where all nodes are connected to their 4 immediate neighbors only. 
When this grid is uniformly weighted, the 2D DCT is shown to be its GFT, and provides an optimal decorrelation of block data modeled by a 2D Gaussian Markov model \cite{zhang2013analyzing}. In this paper, we focus on grids with nearly regular topologies (e.g., all internal nodes have the same number of neighbors) or particular symmetry properties.

%Here, we refer to graphs with $N^2$ nodes as grids, but do not impose any constraint on their graph topology to have more flexibility in designing the transforms. 

\begin{table}[t]
\centering
\caption{Types of symmetric $N\times N$ grids and their corresponding involutions. Indices $k$ and $l$ represent vertical and horizontal coordinates of grid nodes, as in Fig.~\ref{fig:grid_coords}.}
\label{tab:sym_grid}
\begin{tabular}{|l|l|}
\hhline{|=|=|}
Symmetry type & Involution \\
\hhline{|=|=|}
Centrosymmetry & $\phi((k,l))=(N+1-l,N+1-k)$ \\
\hline
UD-symmetry & $\phi((k,l))=(N+1-k,l)$ \\
\hline
LR-symmetry & $\phi((k,l))=(k,N+1-l)$ \\
\hline
Diagonal symmetry & $\phi((k,l))=(l,k)$ \\
\hline
Anti-diagonal symmetry & $\phi((k,l))=(N+1-l,N+1-k)$ \\
\hhline{|=|=|}
\end{tabular}
\end{table}

GFTs on grids with arbitrary weights correspond to 2D non-separable transforms for pixel blocks, which can achieve a significant compression gain over the DCT \cite{arrufat2014nonseparable}.
In our recent work \cite{lu2017fast}, we proposed speedup techniques for 2D grid-based transforms based on various types of grid symmetries: 1)~centrosymmetry, 2)~up down (UD-) symmetry, 3)~left right (LR-) symmetry, 4)~diagonal symmetry, 5)~anti-diagonal, and 6)~grids with multiple symmetry properties. These grid symmetries are defined based on different axes or point of symmetry, as shown in Fig.~\ref{fig:grid_coords}. While \cite{lu2017fast} applies different node re-ordering rules for different types of symmetric grid, in this work we can simply describe grid symmetries by the involutions shown in Table~\ref{tab:sym_grid}, to provide a more straightforward derivation of fast GFTs on symmetric grids. For application, it has been shown in \cite{gnutti2018symmetry} that GFTs with exact or partial symmetry properties can enhance compression efficiency with respect to the DCT.

\begin{figure*}[th]
\centering
\subfigure[]{
\includegraphics[width=.44\textwidth,valign=b]{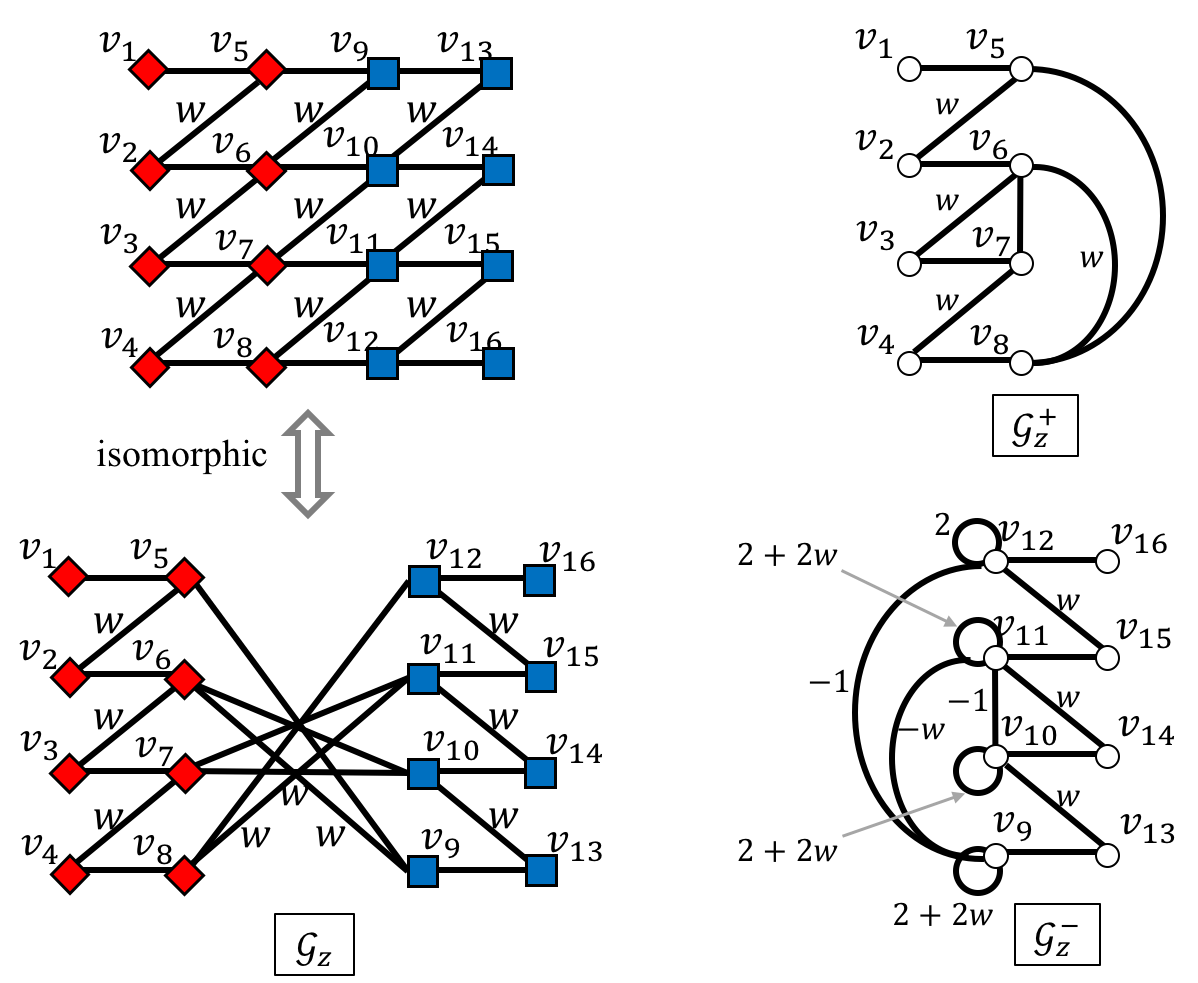}} \hspace{.6cm}
\subfigure[]{
\includegraphics[width=.34\textwidth,valign=b]{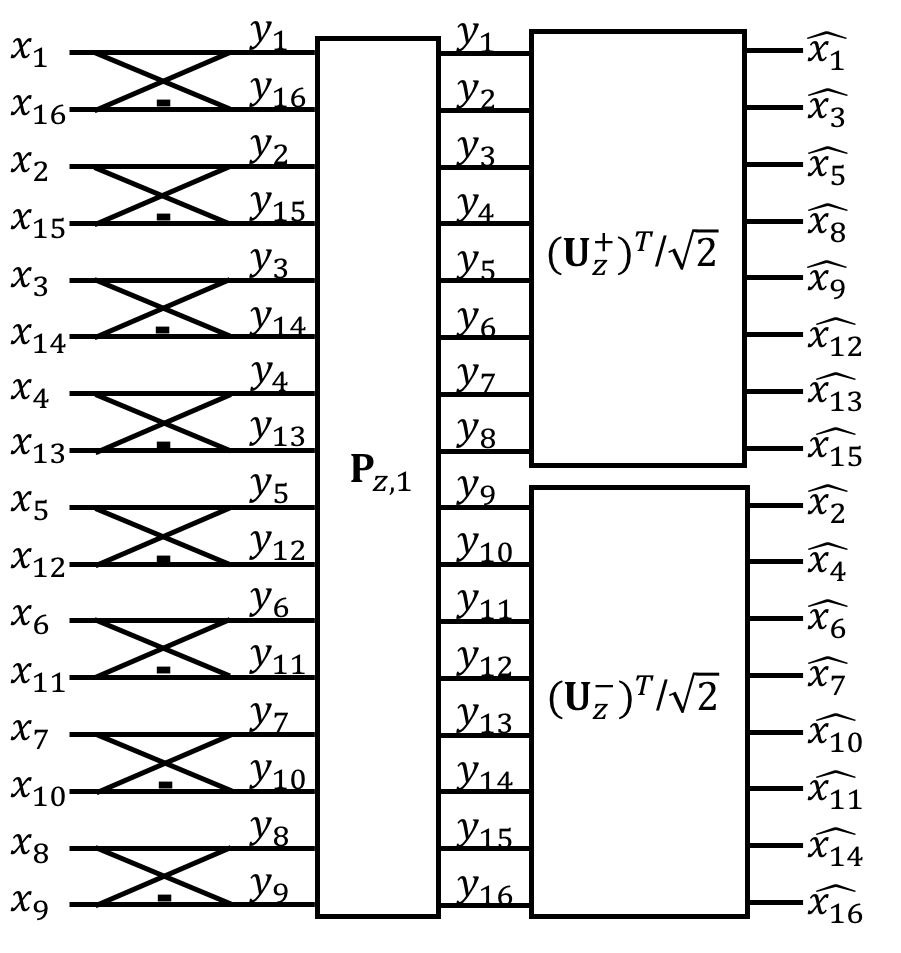}}
\caption{The 4$\times$4 z-shaped grid: (a) Graph decomposition. (b) The associated fast GFT diagram. Red diamonds and blue squares represent those nodes in $\Vc_X$ and $\Vc_Y$, respectively, for the \emph{next} stage of decomposition. Unlabeled edges have weights 1, and $\Pm_{z,1}$ is a permutation operation.}
\label{fig:z4x4}
\end{figure*}

We provide two examples of fast GFTs on symmetric grids. The first example is shown in Fig.~\ref{fig:grid_bidiag}, with a 4$\times$4 grid $\Gc_b$ that is bi-diagonally symmetric (symmetric around both diagonals). We can first decompose $\Gc_b$ based on the diagonal symmetry into $\Gc_b^+$ and $\Gc_b^-$. Then, we observe that the symmetry around the anti-diagonal remains in $\Gc_b^+$ and $\Gc_b^-$, so further decomposition can be applied. As a result, the overall GFT has two butterfly stages of Haar units, and can be implemented using 4 sub-GFTs with length 6, 4, 4, and 2, as in Fig.~\ref{fig:grid_bidiag}. %The overall number of multiplications required is 80, as opposed to 256 in a general 4$\times$4 non-separable transform.
For the second example, we consider a grid graph in the coding framework proposed in \cite{rotondo2015designing}. This grid, which we refer to as \emph{z-shaped grid}, is 4-connected grid with horizontal and anti-diagonal edges, as shown in the top-left of Fig.~\ref{fig:z4x4}(a).
%, where image blocks are classified into $K$ classes, each with an associated GFT to be applied. These GFTs are derived from graph templates including 1)~4-connected grids with horizontal and vertical edges, 2)~4-connected grids with horizontal and anti-diagonal edges as shown in the top-left of Fig.~\ref{fig:z4x4}(a), which we refer to as \emph{z-shaped grid}, and 3)~rotated and flipped versions of the z-shaped grid. Those templates are weighted graphs with the constraint that all edges with the same orientation have a common weight. Without loss of generality, all GFTs from graph templates within types 2) or 3) can be characterized by the GFT of a z-shaped grid with horizontal weights 1 and anti-diagonal weights $w$. 
We denote this grid as $\Gc_z$ and derive its fast GFT in Fig.~\ref{fig:z4x4} based on the centrosymmetry of the grid, characterized by the involution $\phi(i)=N+1-i$. Note that, if we flip the nodes $v_9$ to $v_{16}$ up to down, then $\Gc_z$ becomes a left-right symmetric grid, based on which we can derive $\Gc_z^+$ and $\Gc_z^-$ as in Fig.~\ref{fig:z4x4}(a). The derived fast GFT diagram in Fig.~\ref{fig:z4x4}(b) can thus provide a computational speedup for the coding framework in \cite{rotondo2015designing}.

\subsection{Skeletal Graphs}
\label{subsec:example_skeletal}

\begin{figure*}[th]
%\vspace{-.5cm}
\centering
\subfigure[]{
\includegraphics[width=.62\textwidth,valign=c]{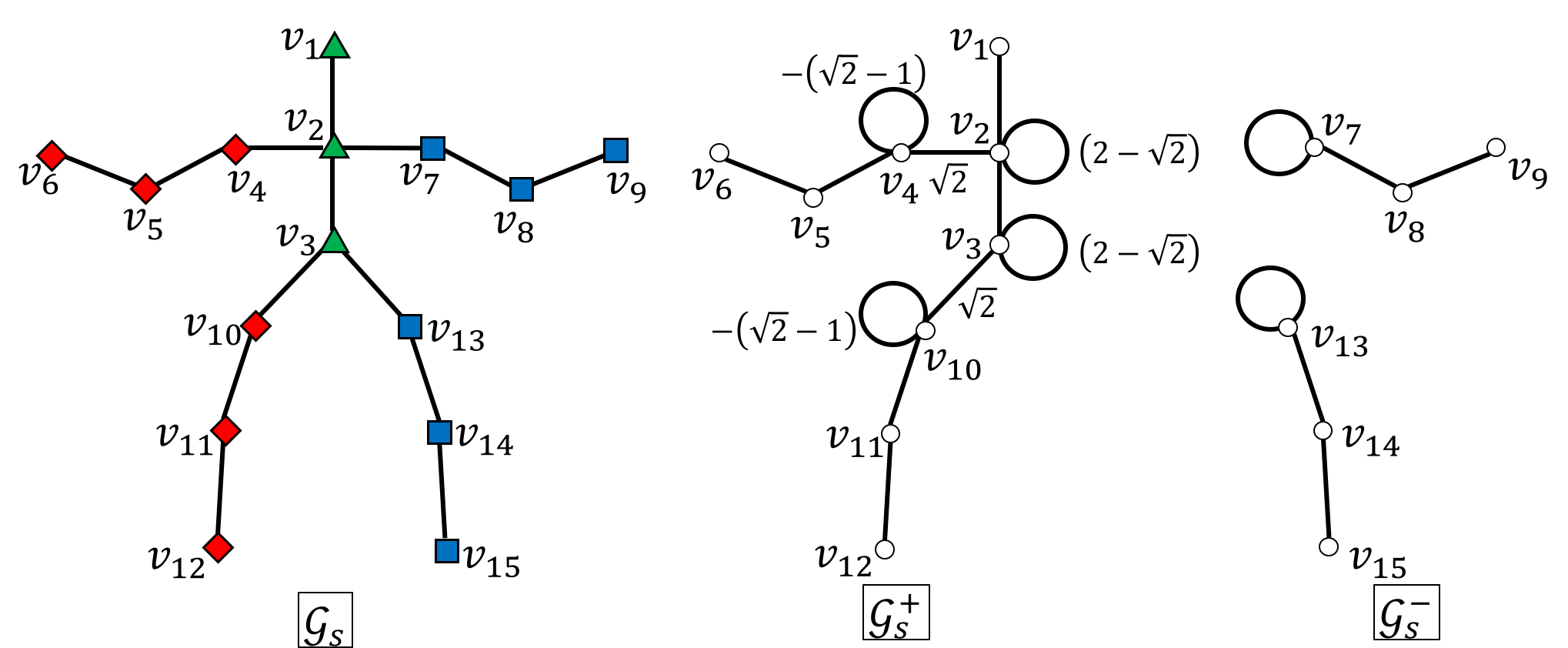}}%
\subfigure[]{
\includegraphics[width=.33\textwidth,valign=c]{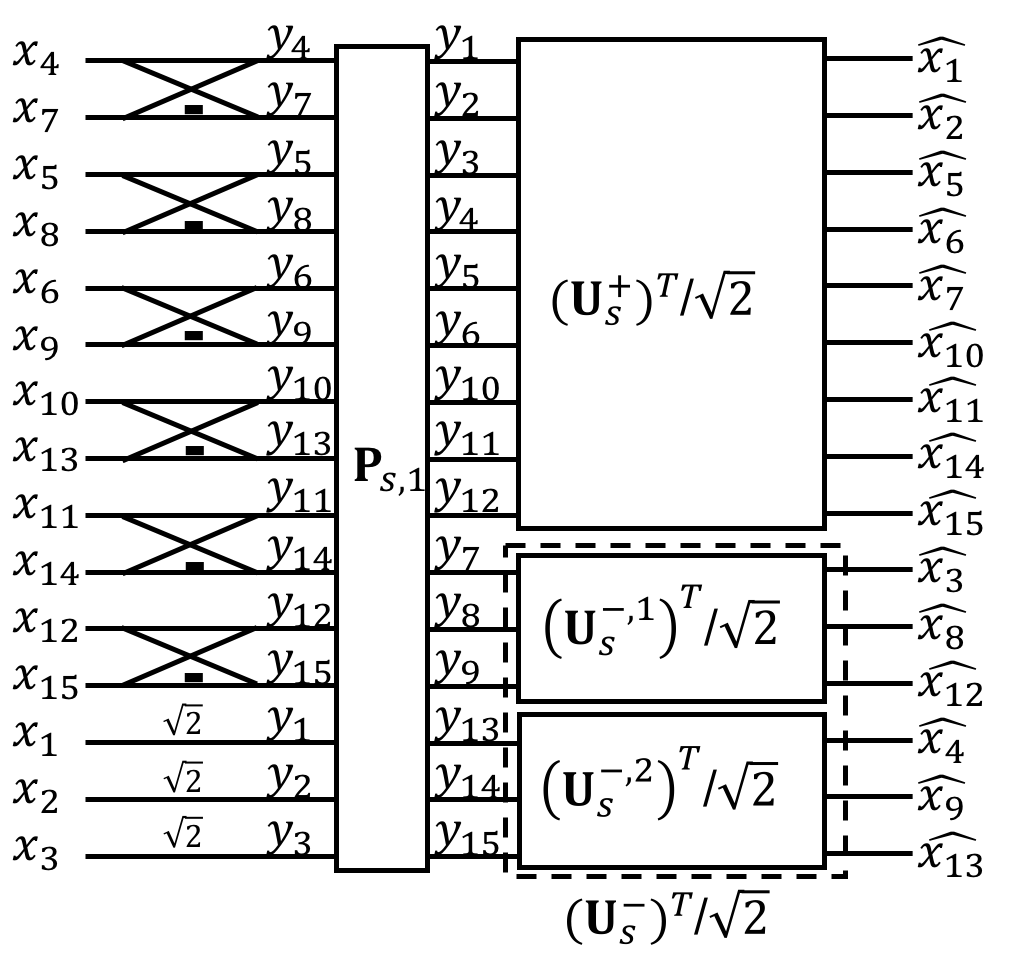}}
\caption{The 15-node skeletal graph: (a) Graph decomposition. (b) The associated fast GFT diagram. Red diamonds, green triangles, and blue squares represent those nodes in $\Vc_X$, $\Vc_Z$, and $\Vc_Y$, respectively, for the \emph{next} stage of decomposition. $\Um_s^{-,1}$ and $\Um_s^{-,2}$ are the GFTs corresponding to two connected components of $\Gc_s^-$, respectively. Unlabeled edges have weights 1, and $\Pm_{s,1}$ is a permutation operation.}
\label{fig:Gsk}
\end{figure*}

In human action analysis, human body can be represented by a hierarchy of joints that are connected with bones. Human motion data can be obtained by two different types of techniques. First, \emph{motion capture} is a process of directly extracting human movements from wearable devices, such as reflective markers attached near each joint. Second, using current methods such as OpenPose \cite{cao2018openpose}, skeletons can be obtained from images or videos in real time. With both techniques, the output of the system contains 3D coordinates of human joints. Then, we can consider the human skeleton as a graph, and motion data on the skeleton as graph signals. Further signal processing techniques can be applied to those signals to perform tasks such as classification and segmentation. 

The work \cite{kao2014graph-based} has demonstrated that the GFT basis of the skeletal graph has localization properties useful in characterizing human motion. For example, the second GFT basis function has positive entries on joints in the upper body, and negative entries on those in the lower body. Thus, the resulting GFT coefficients can provide a discriminating power between different human actions. 

Typical skeletal graphs are symmetric by construction, so a fast GFT can be obtained, as shown in Fig.~\ref{fig:Gsk}, where a 15-node skeleton is considered. Such a fast GFT on skeletal graph can speed up the feature extraction procedure for further action classification tasks. 
Note that the butterfly stage is also available for skeletal graphs with non-uniform weights or different topologies, as long as the desired symmetry properties hold. 
%A separate paper \cite{construction_skeleton} will be devoted to the construction of skeletal graphs and detailed interpretations of their GFT coefficients for human action analysis. 

\subsection{Search of Symmetries in General Graphs}
\label{subsec:search}
In the previous examples, symmetry properties of graphs can be easily identified by inspection. However, in general, and particularly for denser graphs, desired symmetry properties may not be straightforward to identify, or may not even exist. To design fast GFTs for graphs beyond the previous examples, an algorithm for searching a valid involution would thus be useful.

The number of involutions on $n$ elements is \cite[Sec. 5.1.4]{knuth1998art}
\[
  T(n)=\sum_{k=0}^{\lfloor n/2 \rfloor} \frac{n!}{2^k(n-2k)!k!} \sim \left(\frac{n}{e}\right)^{n/2}\frac{e^{\sqrt{n}}}{(4e)^{1/4}},
\]
which asymptotically grows faster than a polynomial in $n$. This means that an exhaustive search of valid involutions among $T(n)$ possible candidates is a combinatorial problem. Here, we provide two methods to reduce the complexity of this search. More detailed illustrations and implementations of these methods can be found in \cite{lu2019search}.
% \begin{itemize}
%     \item {\bf 
\subsubsection{Pruning based on the degree list} 
Note that if $\Gc$ is $\phi$-symmetric, then the degrees of nodes $i$ and $\phi(i)$ must be equal for every $i$. This necessary condition for $\phi$-symmetry allows us to prune the involution search. In particular, we can compute the list of degrees first, then skip searching those involutions $\varphi$ with different degrees on nodes $i$ and $\varphi(i)$ for some $i$. For graphs with many distinct weight values, we tend to have many distinct node degrees, and thus the number of involutions that need to be searched can be significantly reduced. 
\subsubsection{Searching of identical tree branches} 
Trees (i.e., graphs with no cycles) are connected graphs that have the smallest number of edges. This sparsity property implies that symmetry on trees can be characterized by pairs of identical subtrees (i.e., branches) whose roots are common or adjacent. For example, in Fig.~\ref{fig:Gsk}, the two arms in the skeletal graph are identical branches that share a common root, and so are the two legs. 
Based on an algorithm proposed in \cite{flouri2013optimal}, we provide in \cite{lu2019search} an algorithm with $\Oc(n\log n)$ complexity that, for any given tree $\Gc$, finds all involutions $\phi$ such that $\Gc$ is $\phi$-symmetric.
%\end{itemize}

\subsection{Complexity Analysis}
In general, for a length-$n$ fast GFT with a layer of Haar units based on involution $\phi$, the number of multiplications is $p_\phi^2+(n-p_\phi)^2$. This number is minimized to $n^2/2$ when $p_\phi=n/2$. This means that, in the best case scenario with one layer, the overall complexity is reduced by half, and the order of magnitude  remains $\Oc(n^2)$. 
% The complexity in asymptotic notation may be reduced beyond $\Oc(n^2)$ only for some specific graphs, where symmetries in both $\Gc^+$ and $\Gc^-$ are always available. For example, it can be shown that a uniformly weighted complete graph with $n=2^k$ nodes yields $k$ layers of Haar units, leading to an overall complexity of $\Oc(n)$.

\section{Experimental Results}
\label{sec:exp}

In this section, we provide theoretical complexity analysis with the numbers of operations as well as experiments for empirical computation complexities of the fast GFTs. We have implemented several fast GFTs in C to simulate an environment closer to hardware. The source code for the experiments is available at \cite{lu2019search}. 

\begingroup
\renewcommand*{\arraystretch}{1.4}
\begin{table}
\centering
\caption{Speed performance of proposed fast GFTs. The baseline for the runtime reduction rates is the matrix GFT implementation.}
\label{tab:runtime}
\begin{tabular}{|l|c|cc|c|}
\hhline{|=|=|==|=|}
\multirow{2}{*}{Topology} & \multirow{2}{*}{$n$} & \multicolumn{2}{c|}{Number of Operations} & Runtime \\
& & Matrix ($\pm$/$*$) & Fast ($\pm$/$*$) & Reduction \\
\hhline{|=|=|==|=|}
% \multirow{2}{*}{Star graph} & 10 & 90$/$100 & 20$/$22 & 68.1\% \\
% & 100 & 9900$/$10000 & 208$/$137 & 97.5\% \\
% \hline
\multirow{2}{*}{Cycle} & 12 & 132$/$144 & 44$/$30 & 52.7\% \\
& 80 & 6320$/$6400 & 1224$/$1078 & 79.7\% \\
\hline
\multirow{2}{*}{6-conn. grid} & 16 & 240$/$256 & 80$/$80 & 53.7\% \\
& 64 & 4032$/$4096 & 1104$/$1072 & 68.5\% \\
\hline
\multirow{2}{*}{Z-shaped grid} & 16 & 240$/$256 & 128$/$112 & 41.5\% \\
& 64 & 4032$/$4096 & 2048$/$2048 & 45.0\% \\
\hline
% \multirow{2}{*}{4-conn. grid} & 16 & 240$/$256 & 68$/$36 & \% \\
% & 64 & 4032$/$4096 & 532$/$364 & \% \\
% \hline
\multirow{2}{*}{Skeleton} & 15 & 210$/$225 & 96$/$102 & 45.5\% \\
& 25 & 600$/$625 & 272$/$282 & 47.5\% \\
\hhline{|=|=|==|=|}
\end{tabular}
\end{table}
\endgroup

\subsection{Comparison with Matrix GFT}
\label{subsec:exp_1}
In the first experiment, we include the GFTs of several different graph topologies: the cycle graph with unit weights, the bi-diagonally symmetric 6-connected grid (as in Fig.~\ref{fig:grid_bidiag}, with $a=0.5$), the z-shaped grid with $w=2$, and the skeletal graph. For each graph we implement two fast GFTs with different sizes, and compare the runtime between the matrix GFT implementation and the fast GFT with butterfly stages. We include those GFTs in Figs.~\ref{fig:cycle12} to \ref{fig:Gsk}, together with larger graphs with the same topology types: the 80-node cycle graph, the 8$\times$8 bi-diagonally symmetric 4-connected grid, the 8$\times$8 z-shaped grid, and the 25-node skeletal graph used in \cite{shahroudy2016ntu}. Detailed design of their fast GFTs can be extended from the examples in Figs.~\ref{fig:cycle12} to \ref{fig:Gsk}. For each GFT, we generate 20000 graph signals with a proper length, whose entries are i.i.d. uniform random variables with range $[0,1]$. Then, we compute the percentage of runtime reduction for the symmetry-based fast GFT compared to the GFT realized by a single $n\times n$ matrix multiplication. 

In Table \ref{tab:runtime}, we show, for each GFT, the numbers of additions (including subtractions), multiplications, and the empirical computation time reduction rate compared to matrix GFT in C implementation. We see that the fast GFT on skeletal graph in Fig.~\ref{fig:Gsk} with one butterfly stage leads to 45.5\% speed improvement, and that on z-shaped grid in Fig.~\ref{fig:z4x4} gives around 41.5\% runtime saving. Fast GFTs on cycle graphs and 6-connected grids that have multiple butterfly stages yield higher runtime reduction rates. From those results in Table \ref{tab:runtime}, we can see that the butterfly stages obtained from our proposed method lead to a significant speedup, and can be useful if the transform is required to be performed many times, and in a low-level or hardware implementation.

\begin{figure}
    \centering
    \subfigure[8$\times$8 bi-diagonally symmetric 6 connected grid $\Gc_b$]{
    \includegraphics[width=.44\textwidth]{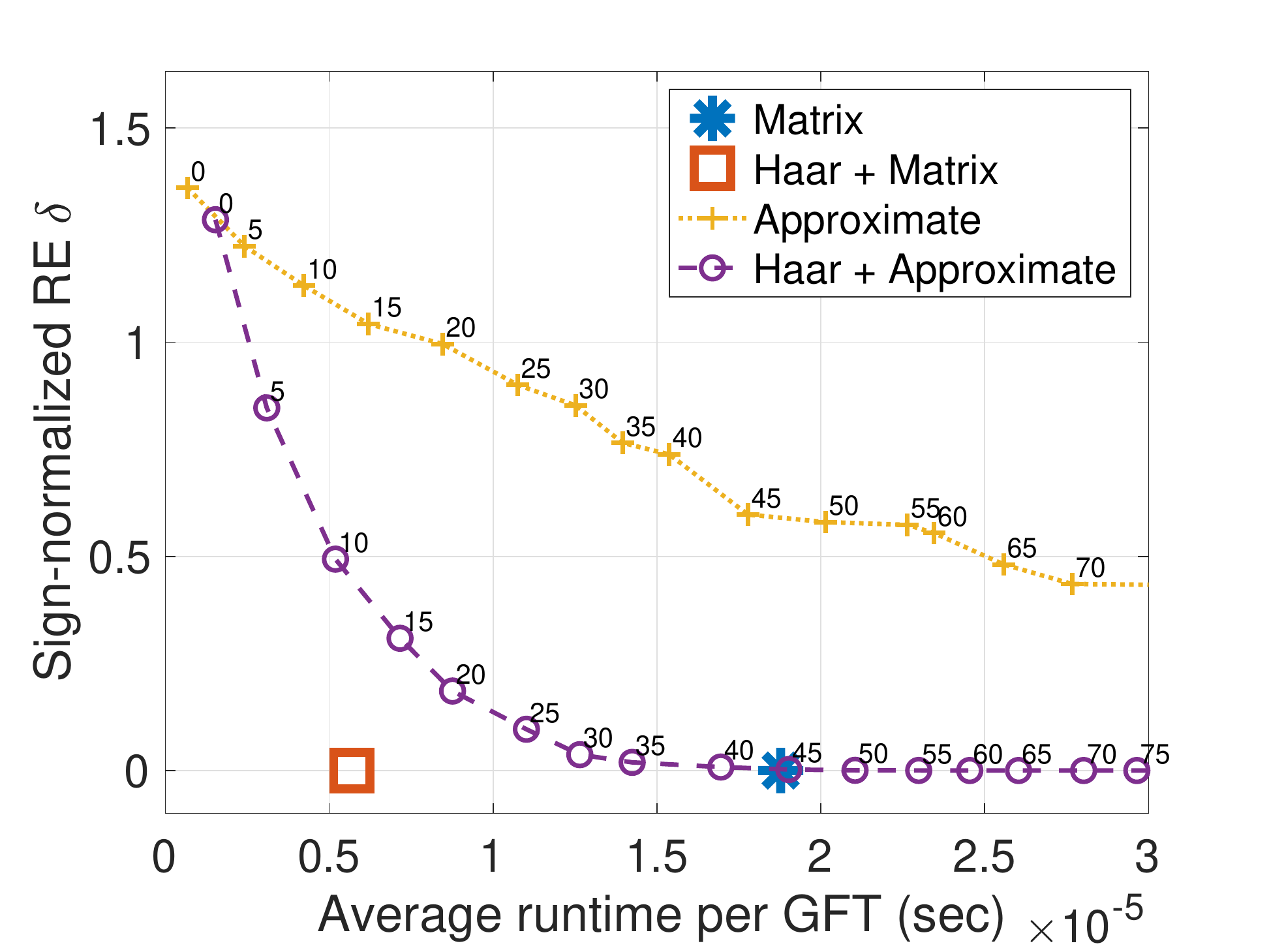}}
    \subfigure[8$\times$8 z-shaped grid $\Gc_z$]{
    \includegraphics[width=.44\textwidth]{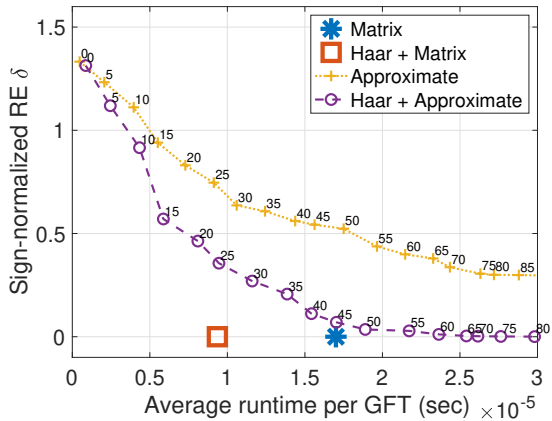}}
    \caption{Runtime versus sign-normalized relative error $\delta$ for different GFT implementations on different graphs. The numbers labeled alongside the markers indicate the associated numbers of Givens rotation layers.}
    \label{fig:runtime_vs_re}
\end{figure}
\begin{figure}
    \centering
    \subfigure[8$\times$8 bi-diagonally symmetric 6 connected grid $\Gc_b$]{
    \includegraphics[width=.44\textwidth]{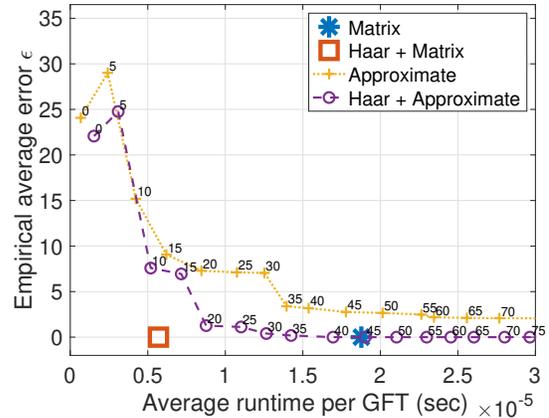}}
    \subfigure[8$\times$8 z-shaped grid $\Gc_z$]{
    \includegraphics[width=.44\textwidth]{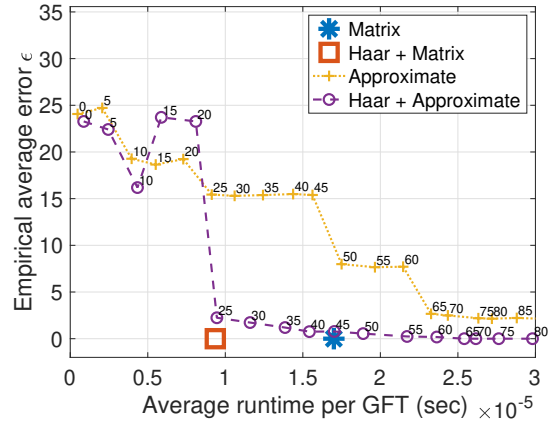}}
    \caption{Runtime versus empirical average error $\epsilon$ for different GFT implementations on different graphs. The numbers labeled alongside the markers indicate the associated numbers of Givens rotation layers.}
    \label{fig:runtime_vs_err}
\end{figure}

\subsection{Comparison with Approximate Fast GFTs}
\label{subsec:exp_2}

In the second experiment, we compare our proposed method with an existing fast GFT approach \cite{magoarou2018approximate} on graphs with symmetry properties. We consider two graphs for this experiment: the 8$\times$8 bi-diagonally symmetric grid with $a=0.5$, and the 8$\times$8 z-shaped grid with $w=2$. Note that, when the desired symmetry property is available, existing methods can be incorporated into the symmetry-based fast GFT scheme to speed up the computation of sub-GFTs such as $\Um^+$ and $\Um^-$. Thus, we can compare the following four GFT implementations: 
\begin{enumerate}
    \item \emph{Matrix GFT}: an $n\times n$ matrix multiplication.
    \item \emph{Haar + matrix GFT}: symmetry-based fast GFT using Haar units, as shown in Figs.~\ref{fig:grid_bidiag}(b) and \ref{fig:z4x4}(b), where the sub-GFTs are implemented by full matrix multiplications.
    \item \emph{Approximate GFT \cite{magoarou2018approximate}}: fast GFT using layers of Givens rotations found by the parallel truncated Jacobi algorithm--a greedy-based algorithm that progressively approximate $\hat{\Um}^\top\Lm\hat{\Um}$ to a diagonal matrix. The resulting GFT can be implemented using the schematic diagram as in Fig.~\ref{fig:layeredgivens}.
    \item \emph{Haar + approximate GFT}: symmetry-based fast GFT with sub-GFTs implemented by approximate GFTs.
\end{enumerate}
Let the GFT matrix associated to a GFT implementation be $\hat{\Um}$, and the true GFT matrix be $\Um$. We define two error metrics as follows. 
\begin{enumerate}
\item \emph{Sign-normalized relative error (RE):} we consider the relative error between two $n\times n$ orthogonal matrices,
\begin{equation}
\label{eq:re}
\text{RE}(\hat{\Um},\Um)=\frac{\|\Um-\hat{\Um}\|_F}{\|\Um\|_F} = \frac{\|\hat{\Um}^\top\Um-\Id\|_F}{\sqrt{n}}. 
\end{equation}
Note that if $\hat{\Um}=-\Um$, the RE will be large although they share a common eigen-structure. To avoid this sign ambiguity, we modify \eqref{eq:re} by taking absolute values elementwise on $\hat{\Um}^\top\Um$:
\[ \delta(\hat{\Um},\Um):=\frac{1}{\sqrt{n}}\| |\hat{\Um}^\top\Um|-\Id \|_F. \]
\item \emph{Empirical average error:} let $\Xc=\{\xv_1,\dots,\xv_M\}$ be the set of input signals, we define
\[
  \epsilon(\hat{\Um},\Um,\Xc):=\frac{1}{M}\sum_{i=1}^M \sum_{j=1}^n \left(|\uv_j^\top\xv_i|-|\hat{\uv}_j^\top\xv_i|\right)^2,
\]
where $\uv_j^\top\xv_i$ and $\hat{\uv}_j^\top\xv_i$ are the $j$-th true and approximate GFT coefficients of $\xv_i$. The absolute values are used to avoid the sign ambiguity. 
\end{enumerate}
For both graphs considered in this experiment, the eigenvalues of the Laplacians are all distinct, so the GFT bases have no rotation ambiguity.

We apply the method in \cite{magoarou2018approximate} to obtain the parameters (angle and node pairings for Givens rotations) approximate GFTs, then implement the resulting fast algorithms in C, with different numbers of layers $J\in\{0,5,10,\dots\}$. We use $M=20000$ random samples as in Sec.~\ref{subsec:exp_1}, and obtain the error metrics, $\delta$ and $\epsilon$, for each GFT implementation. 

The runtime versus sign-normalized RE, and versus empirical average error are shown in Figs.~\ref{fig:runtime_vs_re} and \ref{fig:runtime_vs_err}, respectively. We note that, first, the RE drops more steadily than the empirical error when the number of layers increases. This is related to the order of GFT basis functions. When more layers of Givens rotations are introduced, more GFT basis functions will be ordered correctly (i.e., smaller error in the final permutation operation $\Pim_{J+1}$ in Fig.~\ref{fig:layeredgivens}). Indeed, we observe that when the number of correctly ordered GFT coefficients increases, the decrease of the empirical error is usually more significant than that of the relative error. The second observation is that for the two graphs with $n=64$ nodes, the approximate GFTs typically takes more than 20 layers to yield a sufficiently accurate GFT in terms of both error metrics. However, when more than 20 layers are used, the computation complexity becomes comparable or higher than Haar-matrix GFT, which provides exact GFT coefficients. Finally, we see that in both Figs.~\ref{fig:runtime_vs_re} and \ref{fig:runtime_vs_err}, the error of Haar + approximate GFT drops faster than that of approximate GFT. This means that by applying the symmetry property, we can obtain a significantly higher convergence rate for the approximation. This is a reasonable consequence, as our divide-and-conquer method reduces the dimension of the parameter estimation problem for \cite{magoarou2018approximate}. 

\section{Conclusion}
\label{sec:conclusion}
In this paper, we have explored the relationship between the graph topology and properties in the corresponding GFT for fast GFT algorithm based on butterfly stages. We focus particularly on a component of the butterfly stage called Haar unit, and discuss the conditions for a stage of Haar units to be available in the GFT implementation. We have shown that a graph has a right butterfly stage with Haar units if it is k-regular bipartite. On the other hand, a left butterfly stage is available if the graph has certain symmetry properties. We have formally defined the relevant graph symmetry based on involution, i.e., pairing function of nodes. Then, we have proposed an approach, where once a graph symmetry is identified, we can decompose a graph $\Gc$ into two smaller graphs, $\Gc^+$ and $\Gc^-$, whose GFTs corresponds to the two parallel sub-transforms after a butterfly stage of Haar units. Again, from $\Gc^+$ and $\Gc^-$ we can explore subsequent butterfly stages if any desired symmetry property holds in them. Thus, this method enables us to explore butterflies stage by stage. 

The desired symmetry properties typically arise in graphs that are nearly regular, symmetric by construction, or uniformly weighted. We have discussed several classes of those graphs: bipartite graphs, graphs with 2-sparse eigenvectors such as star and complete graphs, symmetric line and grid graphs, cycle graphs, and skeletal graphs. Relevant applications of those GFTs include video compression and human action analysis. Finally, we implement the fast GFT algorithms in C and compute the runtime saving for several graphs. The experiment results show that our method provides a significant computation time reduction compared to the GFT computed by matrix multiplication. It also outperforms existing fast approximate GFT approaches in terms of both complexity and accuracy for graphs with desired symmetry properties.

\appendix
\subsection{Proof of Theorem \ref{thm:decompose}}
\label{app:d2}
In the following, we will repeatedly use \eqref{eq:sw_from_l} and \eqref{eq:l_from_sw}. Also recall that in \eqref{eq:Lpm},
\begin{align}
\label{eq:Lmp}
  \Lm^+ &= 
  \begin{pmatrix}
  \Lm_{XX}+\Lm_{XY}\Jm & \sqrt{2}\Lm_{XZ} \\ \sqrt{2}\Lm_{XZ}^\top & \Lm_{ZZ}
  \end{pmatrix}, \\
\label{eq:Lmm}
  \Lm^- &= \Lm_{YY}-\Jm\Lm_{XY}.
\end{align}
From the block partition structure \eqref{eq:L_123}, we also have 
\begin{align*}
  & (\Lm_{XX})_{i,j}=l_{i,j},\quad (\Lm_{XZ})_{i,j}=l_{i,p+j},\quad (\Lm_{XY})_{i,j}=l_{i,n-p+j},\quad \\
  & (\Lm_{ZZ})_{i,j}=l_{p+i,p+j}, \quad (\Lm_{YY})_{i,j}=l_{n-p+i,n-p+j}, \\
  & (\Jm\Lm_{XY})_{i,j}=l_{p+1-i,n-p+j}, \quad (\Lm_{XY}\Jm)_{i,j}=l_{i,n+1-j}.
\end{align*}
Such changes of indices will be used in the derivations below.
%\begin{itemize}
\subsubsection{Edges of $\Gc^+$} 
With $i,j\in\Vc^+=\{1,\dots,n-p\}$ and $i\neq j$, we discuss three different 
cases separately, all based on \eqref{eq:Lmp}. If $i,j\in\Vc_X=\{1,\dots,p\}$, then
  \[
    w_{i,j}^+=-\left(\Lm_{XX}+\Lm_{XY}\Jm\right)_{i,j}
    =-(l_{i,j}+l_{i,n+1-j})=w_{i,j}+w_{i,n+1-j}. 
  \]
If $i\in\Vc_X=\{1,\dots,p\}$ and $j\in\Vc_Z=\{p+1,\dots,n-p\}$, then 
\[
  w_{i,j}^+=-\left(\sqrt{2}\Lm_{XZ}\right)_{i,j-p} = -\sqrt{2}l_{i,j} = \sqrt{2}w_{i,j},
\]
and the same holds for $i\in\Vc_Z$ and $j\in\Vc_X$. If $i,j\in\Vc_Z=\{p+1,\dots,n-p\}$, then 
\[
  w_{i,j}^+ = -\left(\Lm_{ZZ}\right)_{i-p,j-p} = -l_{i,j}=w_{i,j}.
\]
%By \eqref{eq:Lmp}, those edges within nodes $\Vc_X=\{1,\dots,p\}$ correspond to the off-diagonal 
%elements of $\Lm_{XX}+\Lm_{XY}\Jm$, so they can be obtained in the same way as in Remark 
% \ref{rem:construct}. Those edges within $\Vm_m$ are associated to $\Lm_{ZZ}$, so they are the same as in $\Gc$. 
%The weights of edges across $\Vc_X$ and $\Vc_Z$ are scaled by $\sqrt{2}$ as they are in $\Gc$.
\subsubsection{Self-loops of $\Gc^+$} 
To express $s_i^+$, we discuss the cases with $i\in\Vc_X$ and 
$i\in\Vc_Z$ separately. If $i\in\Vc_X=\{1,\dots,p\}$, then from \eqref{eq:Lmp} we have 
\begin{align*}
  s_i^+ &=\sum_{j=1}^p \left(\Lm_{XX}+\Lm_{XY}\Jm\right)_{i,j}
    +\sum_{j=1}^{n-2p}\left(\sqrt{2}\Lm_{XZ}\right)_{i,j} \\
  &=\sum_{j=1}^p(l_{i,j}+l_{i,n+1-j}) + \sqrt{2}\sum_{j=1}^{n-2p}l_{i,p+j} \\
  &=\underbrace{s_i + \sum_{\substack{j=1 \\ j\neq i}}^n w_{i,j}}_{l_{i,i}}
  + \underbrace{\left(-\sum_{\substack{j=1 \\ j\neq i}}^p w_{i,j}\right)}_{l_{i,j} \text{ with }i\neq j}
  - \sum_{j=1}^p w_{i,n+1-j} - \sqrt{2}\sum_{j=1}^{n-2p}w_{i,p+j} \\
  %&= s_i +\sum_{j=p+1}^n w_{i,j} - \sum_{j=n-p+1}^n w_{i,j} - \sqrt{2}\sum_{j=p+1}^{n-p} w_{i,j} \\
  &= s_i - (\sqrt{2}-1)\sum_{j=p+1}^{n-p} w_{i,j}.
\end{align*}
If $i\in\Vc_Z=\{p+1,\dots,n-p\}$, then \eqref{eq:Lmp} gives
\begin{align*}
  s_i^+ &=\sum_{j=1}^{n-2p} \left(\Lm_{ZZ}\right)_{i-p,j}
    +\sum_{j=1}^{p}\left(\sqrt{2}\Lm_{XZ}\right)_{i-p,j} \\
  &= \sum_{j=1}^{n-2p} l_{i,p+j} + \sqrt{2}\sum_{j=1}^p l_{j,i} \\
  &= \underbrace{s_i + \sum_{\substack{j=1 \\ j\neq i}}^n w_{i,j}}_{l_{i,i}} 
  + \underbrace{\left(-\sum_{\substack{j=p+1 \\ j\neq i}}^{n-p} w_{i,j}\right)}_{l_{i,j}\text{ with }i\neq j} - \sqrt{2}\sum_{j=1}^p w_{i,j} \\
  &= s_i + \sum_{j=1}^p w_{i,j} + \sum_{j=n-p+1}^n w_{i,j} -\sqrt{2}\sum_{j=1}^p w_{i,j} \\
  &= s_i + (2-\sqrt{2})\sum_{j=1}^p w_{i,j}
\end{align*}

\subsubsection{Edges of $\Gc^-$} 
With $i,j\in\Vc_Y=\{n-p+1,\dots,n\}$ and $i\neq j$, from \eqref{eq:Lmm} we have 
  \begin{align*}
    w^-_{i,j}&=-\left(\Lm_{YY}-\Jm\Lm_{XY}\right)_{i-n+p,j-n+p}\\
    &=-(l_{i,j}-l_{n+1-i,j})=w_{i,j}-w_{n+1-i,j}. 
  \end{align*}
%Because \eqref{eq:Lmm} has the same form as in Section \ref{sec:decomposition} ($\Lm^-=\Lm_{ZZ}-\Jm\Lm_{XZ}$), 
%its edge weights can be obtained in the same way as in Remark \ref{rem:construct}.
\subsubsection{Self-loops of $\Gc^-$} 
Here, we have $i\in\Vc_Y=\{n-p+1,\dots,n\}$, so, by \eqref{eq:Lmm},
\begin{align*}
  s_i^- &= \sum_{j=1}^p\left(\Lm_{YY}-\Jm\Lm_{XY}\right)_{i-n+p,j}
  = \sum_{j=1}^p(l_{i,n-p+j}-l_{n+1-i,n-p+j}) \\
  &= \underbrace{s_i+\sum_{\substack{j=1 \\ j\neq i}}^n w_{i,j}}_{l_{i,i}}
   + \underbrace{\left(-\sum_{\substack{j=n-p+1 \\ j\neq i}}^n w_{i,j}\right)}_{l_{i,j}\text{ with }i\neq j}
   + \sum_{j=1}^p w_{i,p+1-j} \\
  &= s_i + 2\sum_{j=1}^{p} w_{i,j} + \sum_{j=p+1}^{n-p} w_{i,j}.
\end{align*}
The results derived above apply to the case when $\phi=(n,n-1,\dots,1)$. For any arbitrary $\phi$, we can simply modify the sub-indices based on $\phi$. Thus, the results above can be written as in Lemma \ref{thm:decompose}.
%\end{itemize}

%\section*{Acknowledgment}

\ifCLASSOPTIONcaptionsoff
  \newpage
\fi

\bibliographystyle{IEEEbib}
\bibliography{refs}

% \begin{IEEEbiography}{Keng-Shih Lu}
% \end{IEEEbiography}

% \begin{IEEEbiography}{Antonio Ortega}
% \end{IEEEbiography}

%\newpage


%\vfill


\end{document}